%% file: article.tex
\documentclass[entropy,article,accept,moreauthors,pdftex]{Definitions/mdpi}

\usepackage[T1]{fontenc}
\usepackage[utf8]{inputenc}

\usepackage[labelformat=simple]{subfig}

\usepackage{pgfplots}
\usepackage{tikz}
\usepackage{amsmath}
\usepackage{amssymb}
\usepackage{algorithm}
\usepackage[noend]{algorithmic}
\usepackage{mathtools}

\firstpage{1} 
\pubvolume{1}
\issuenum{1}
\articlenumber{0}
\pubyear{2021}
\copyrightyear{2021}
\externaleditor{Academic Editor: Ernestina Menasalvas}
\datereceived{1 August 2021} 
\dateaccepted{24 August 2021}
\datepublished{} 
\hreflink{https://doi.org/} % If needed use \linebreak

\usetikzlibrary{patterns}
\usetikzlibrary{positioning}
\usetikzlibrary{shapes}
\usetikzlibrary{matrix}
\usetikzlibrary{topaths}
\usepgfplotslibrary{external}

\pgfplotsset{compat = 1.3}

\DeclarePairedDelimiter{\ceil}{\lceil}{\rceil}

\tikzstyle{node} = [circle, draw, minimum size = 0.6 cm]

\DeclareMathOperator{\AU}{AU}
\DeclareMathOperator{\CU}{CU}
\DeclareMathOperator{\cost}{cost}
\DeclareMathOperator{\decide}{decide}
\DeclareMathOperator{\length}{length}
\DeclareMathOperator{\mylabel}{label}
\DeclareMathOperator{\Poisson}{Poisson}
\DeclareMathOperator{\COST}{COST}
\DeclareMathOperator{\pop}{pop}
\DeclareMathOperator{\push}{push}
\DeclareMathOperator{\myrelax}{relax}
\DeclareMathOperator{\swap}{swap}
\DeclareMathOperator{\target}{target}
\DeclareMathOperator{\trace}{trace}
\DeclareMathOperator{\vertex}{vertex}

\newcommand{\twodots}{\mathinner {\ldotp \ldotp}}
\newcommand{\CUO}[2]{[#1\twodots#2]}
\newcommand{\eattrs}[2]{(#1, #2)}
\newcommand{\tn}[4]{(#1, #2, #3, #4)}
\newcommand{\nulledge}{e_\emptyset}

\Title{Towards an Efficient and Exact Algorithm for Dynamic Dedicated
 Path Protection}%Attention AE/ME. The following layout issues have not been checked by the English Editing Department and must be carefully verified by the AE/Layout Department: All callout issues, bold usage of callouts, and references to callouts in the text. Correct callout usage in figures. Figure and Table layout issues. Footnote formatting and Glossaries have not been checked. En dash usage for negative values, en dash usage to indicate relationships, en dash usage to indicate bonds (especially in chemistry). The English Editing Department is not responsible for correct italic usage for genes, proteins and technical terminology. This responsibility belongs to the authors. The following are also not checked: spacing between numbers and units of measurement, ratios, en dashes for ranges, date and time formats, punctuation in equation lines, and less than/more than spacing (< >). Finally, capitalization and layout of titles/headings must be properly checked as well as ensuring 'Eq.' and 'Fig.' are properly spelled out, as these are layout issues.

\TitleCitation{Towards an Efficient and Exact Algorithm for Dynamic Dedicated
 Path Protection}

\Author{Ireneusz Szcześniak $^{1,}$*, Ireneusz Olszewski $^{2}$ and Bożena
 Woźna-Szcześniak $^{3}$}

\AuthorNames{Ireneusz Szcześniak, Ireneusz Olszewski and Bożena
 Woźna-Szcześniak}

\AuthorCitation{Szcześniak, I.; Olszewski, I.; Woźna-Szcześniak, B.}
	
\address{%
$^{1}$ \quad Department of Computer Science, Częstochowa
  University of Technology, 42-200 Częstochowa,
 Poland\\

 $^{2}$ \quad Institute of Telecommunications, UTP University
 of Sciences and Technology, 85-796 Bydgoszcz,
 Poland; irek@utp.edu.pl\\

 $^{3}$ \quad Department of Mathematics and Computer Science, 
 Jan Długosz University, 42-200 Częstochowa,
 Poland; b.wozna@ujd.edu.pl}

\corres{Correspondence: ireneusz.szczesniak@pcz.pl}

\abstract{We present a novel algorithm for dynamic routing with
dedicated path protection which, as the presented simulation results
suggest, can be efficient and exact. We present the algorithm in
the setting of optical networks, but it should be applicable to
other networks, where services have to be protected, and where the
network resources are finite and discrete, e.g., wireless radio or
networks capable of advance resource reservation. To the best of
our knowledge, we are the first to propose an algorithm for this
long-standing fundamental problem, which can be efficient and exact,
as suggested by simulation results. The algorithm can be efficient
because it can solve large problems, and it can be exact because
its results are optimal, as demonstrated and corroborated by
simulations. We offer a worst-case analysis to argue that the
search space is polynomially upper bounded. Network operations,
management, and control require efficient and exact algorithms,
especially now, when greater emphasis is placed on network
performance, reliability, softwarization, agility, and return on
investment. The proposed algorithm uses our generic Dijkstra
algorithm on a search graph generated ``on-the-fly'' based on the
input graph. We corroborated the optimality of the results of the
proposed algorithm with brute-force enumeration for networks up to
15 nodes large. We present the extensive simulation results of
dedicated-path protection with signal modulation constraints for
elastic optical networks of 25, 50, and 100 nodes, and with 160, 320,
and 640 spectrum units. We also compare the bandwidth blocking
probability with the commonly-used edge-exclusion algorithm. We had
48,600~simulation runs with about 41 million searches.}

\keyword{dynamic dedicated path protection; generic Dijkstra
algorithm; elastic optical network; modulation constraints}

\begin{document}

\section{Introduction}
\label{introduction}

% We start with routing, because that's what the article is about.

Optical networks, which are the backbone of communication networks,
need to provide protection for the carried traffic to prevent
large-scale disruptions due to fiber cuts, human errors, hardware
failures, power outages, natural disasters or attacks
\cite{10.1007/s11107-015-0532-0,10.1109/MNET.2015.7340430,10.1007/978-3-319-05227-4}. From among the various ways of
protecting traffic in optical networks, dedicated path protection
(DPP) is the simplest, and the most effective, albeit the most
expensive. In DPP, there are two paths established for a single
demand: the working one, and the protecting one. When the working
path fails, the protecting path delivers the traffic. DPP has been
commonly used and studied for decades.

% WDM networks, EONs.

In a wavelength-division multiplexed (WDM) network, if a client signal
does not fully utilize the fixed spectrum of the assigned wavelength,
the spectrum of the precious erbium window is wasted, a problem
addressed by elastic optical networks (EONs) which divide the spectrum
into fine \emph{frequency slot units} (of, e.g., 12.5 GHz width), or
just \emph{units}, and then allocating \emph{contiguous units} to form
a \emph{slot} tailored to a specific demand
\cite{10.1109/MCOM.2012.6146481}.

% WDM, RWA, RSMA; dynamic vs static.

Routing in WDM networks with the spectrum continuity constraint is
called routing and wavelength assignment (RWA). Routing in EONs with
the spectrum contiguity constraint added is called routing and
spectrum assignment (RSA), and with the signal modulation constraint
added is called routing, modulation, and spectrum assignment (RMSA).
These routing problems can be dynamic or static. In dynamic (aka
online) routing, a single demand is routed in a loaded network, as
opposed to \emph{static} (aka offline) routing, where many demands are
routed in an unloaded network.

% The objectives of routing.

When finding an exact solution for a dynamic routing problem in
optical networks, some path cost is minimized, and the spectrum and
modulation constraints are met. The path cost can be defined in
various ways, e.g., the path length, the number of edges, some signal
quality measure, monetary cost, or a measure related to availability.
The path cost can take into account the cost of traversing not only an
edge, but a vertex, too. In routing with DPP, the cost of a path pair
is minimized, and the spectrum and modulation constraints must be met
for both paths.

% Optimality.

Whether routing along paths of lowest cost leads to optimal network
performance (as measured, for instance, with the bandwidth blocking
probability over a series of established and terminated connections)
is, to the best of our knowledge, an open research problem, which we
do not address in this work. We research the problem of an exact
algorithm, one which finds an optimal (i.e., of lowest cost) solution,
and not of an optimal algorithm. An algorithm optimality could imply
optimal network performance, or optimal computational complexity, and
we address neither of these.

% Contribution in a nutshell: algorithm, simulations, implementation.

Our novel contribution is an algorithm which solves the dynamic
routing problems \emph{with DPP} for WDM networks and EONs without
signal regeneration and spectrum conversion. The algorithm can take
into account various spectrum allocation policies. With extensive
simulations, we demonstrate the computational performance of the
proposed algorithm, which can be polynomial, not exponential. We
corroborate the optimality of the results found for networks up to 15
nodes. Finally, we provide under a liberal license our free and
open-source implementation of the proposed algorithm
\cite{ddppwebsite}.

% Article organization.

The article is organized as follows. In Section \ref{related}, we
review related works, in Section \ref{statement}, we state the research
problem, in Section \ref{algorithm}, we describe the algorithm, and, in
Section \ref{simulations}, we report on the simulation results.
Finally, Section \ref{conclusion} concludes the article.

%%%%%%%%%%%%%%%%%%%%%%%%%%%%%%%%%%%%%%%%%%%%%%%%%%%%%%%%%%%%%%%%%%%%%%%%%%%

\section{Related Works}
\label{related}

% What's the generic Dijkstra algorithm, and how we use it.

The proposed algorithm is based on the \emph{generic} Dijkstra
algorithm recently proposed~\cite{10.1364/JOCN.11.000568}. The
generic Dijkstra algorithm is a generalization of the Dijkstra
algorithm, which takes into account the spectrum continuity and
contiguity constraints by introducing the \emph{incomparability
relation} between solutions. Specifically, we modify the generic
Dijkstra algorithm to work on a search graph, which is built using the
input graph, and represents the possible ways of finding path pairs.
Furthermore, we introduce the incomparability relation between pairs
of paths.

% No other algorithm published before. NP-completeness proofs, yes.

To the best of our knowledge, no efficient and exact algorithm (at
least demonstrated by simulations) for solving the dynamic routing
problem with DPP in optical networks has been published. In
\cite{10.1109/INFCOM.2004.1354524}, the authors offered a proof that
the dynamic RWA with DPP is nondeterministic polynomial time complete
(NP-complete). In Reference \cite{10.1109/ACCESS.2019.2901018}, the authors
offered a proof that the dynamic RMSA with DPP is NP-complete. In
contrast, we propose an algorithm with the efficiency and exactness
demonstrated by simulations, which suggests the problem may be
tractable.

% The routing without DPP.

Dynamic routing without DPP is simpler, but its status seemed unclear.
In Reference \cite{10.1109/TNET.2011.2138717,10.1364/JOCN.6.001115}, the
problem was solved with exponential worst-case time and memory
complexities. However, the dynamic routing problems in EONs can be
solved exactly in polynomial time with the spectrum scan method
\cite{10.1109/CC.2013.6506930}, introduced in Reference 
\cite{10.1117/12.904013}. The exact-routing concept of the spectrum
scan method was introduced earlier for WDM networks in Reference 
\cite{10.1016/S0140-3664(00)00236-X} but was called a
\emph{heuristic} greedy algorithm. That concept was used under the
name of the spectrum window planes \cite{10.1109/JLT.2015.2421506},
and the filtered-graphs algorithm \cite{10.1364/JOCN.11.000568}. In Reference 
\cite{10.1109/GLOCOM.1995.502554}, the authors solved efficiently (in
polynomial time) and exactly the dynamic routing problem in WDM
networks with their interconnected-layered-graph algorithm. That
\emph{exact} algorithm was later improved and applied to EONs in
Reference \cite{10.1364/JOCN.8.000507} but was called \emph{heuristic}.

% Multicriteria shortest path problems.

In a very broad sense, dynamic routing problems are multicriteria
shortest path problems, which, in turn, are multiobjective combinatorial
optimization problems with a set of constraints given to define the
combinatorial structure of the problem \cite{10.1007/s002910000046}.
Whether a specific routing problem is tractable or not depends on the
number and type of criteria (or objective functions) and constraints.
Routing problems are defined for discrete or continuous criteria:
discrete for, e.g., optical networks \cite{10.1109/TNET.2011.2138717}
and networks capable of advance resource reservation
\cite{10.1016/j.comnet.2008.06.016}, continuous for, e.g., networks
with quality-of-service requirements \cite{10.1109/TNET.2004.836112,10.1109/49.536364}. In Reference \cite{10.1007/978-3-642-48782-8_9}, ten
bicriteria shortest path problems were studied, some of them were
proven NP-complete, others were solved in polynomial time with a novel
multilabeling algorithm. That bicriteria multilabeling algorithm was
generalized to any number and type of criteria in Reference 
\cite{10.1016/0377-2217(84)90077-8}, which is now called the
\emph{Martins algorithm}.

% The Martins algorithm.

The Martins algorithm is the basic algorithm for exactly solving any
multicriteria shortest path problem, but with exponential worst-case
memory and time complexities \cite{10.2478/v10006-007-0023-2}. To use
the Martins algorithm for dynamic routing in optical networks, we
could consider available spectrum units as discrete criteria, but that
would lead to exponential worst-case time and memory complexities.
The generic Dijkstra is similar to the Martins algorithm in that it is
also a multilabeling algorithm, but the generic Dijkstra algorithm is
a single criterion shortest path algorithm, where the ordering between
solutions (labels) is partial.

% The classical graph algorithms.

The efficient and exact algorithms for finding a shortest pair of
edge-disjoint paths in a graph are: the Suurballe algorithm
\cite{10.1002/net.3230040204}, the Bhandari algorithm \cite{bhandari},
and any minimum-cost, maximum-flow algorithm (e.g., the successive
shortest path algorithm) with edge capacities set to one
\cite{networkflows}, all of which use the path augmentation technique.
These algorithms cannot be used for solving the stated problem
because they do not consider the spectrum continuity and contiguity
constraints.

% Two well-known approaches.

In our simulations, we also used two well-known algorithms for solving
the problem: the heuristic \emph{edge-exclusion} algorithm, and the
exact \emph{brute-force} algorithm.

% The edge exclusion algorithm.

The edge-exclusion algorithm is a simple and commonly-used algorithm
for finding a pair of edge-disjoint paths: find a shortest path, then
remove from the graph the edges found, and then find a shortest path
again. This heuristic performs quite well, but it often finds suboptimal
solutions, and it can fail even when a solution exists (e.g., for the
so-called trap topology).

% The edge-exclusion algorithm usually uses the KSP algorithm.

The edge-exclusion algorithm usually employs the limited (i.e., with a
limited $K$, e.g., $K = 10$) K-shortest path (KSP) algorithm to find a
shortest path, which is a heuristic algorithm whose blocking
probability depends on the value of $K$. However, the edge-exclusion
algorithm can perform better if an algorithm of lower blocking
probabilities is used. The blocking probabilities of the generic
Dijkstra algorithm can be even twice as low as the blocking
probabilities of the limited KSP algorithm
\cite{10.1109/ONDM.2016.7494087}. Therefore, in the edge-exclusion
algorithm, we used the generic Dijkstra algorithm.

% The brute force algorithm.

The brute-force algorithm enumerates the path pairs using a priority
queue that sorts the pairs in increasing-cost order. After we
retrieve a pair from the queue, we produce new path pairs by extending
one of the paths in the pair with an available edge that was not used
before because the two paths should be edge-disjoint and without
loops. We put a new path pair into the queue, if its paths meet the
spectrum continuity and contiguity constraints. We keep looking for
path pairs until we find one whose paths end at the destination node,
provided we have enough time and memory. We successfully used the
brute-force algorithm only for very small networks (15 nodes), since
this algorithm is very~inefficient.

%%%%%%%%%%%%%%%%%%%%%%%%%%%%%%%%%%%%%%%%%%%%%%%%%%%%%%%%%%%%%%%%%%%%%%%%%%%
\clearpage 
\section{Problem Statement}
\label{statement}

% What we are given.

\noindent{}Given:

\begin{itemize}

\item directed multigraph $G = (V, E)$, where $V = \{v_i\}$ is a set
 of vertexes, and $E = \{e_i\}$ is a set of edges,

\item available units function $\AU(e_i)$, which gives the set of
 available units of edge $e_i$, which do not have to be contiguous,

\item $s$ and $t$ are the source and target vertexes of the demand,

\item a cost function $\cost(p)$, which returns the cost of path $p$,
 
\item a monotonically nondecreasing cost function $\COST(l)$, which
 returns the (real or integer) cost of path pair $l$,

\item a decision function $\decide(p)$ of monotonically increasing
 requirements, which returns true if path $p$ can support the demand,
 otherwise false,

\item the set of all units $\Omega$ on every edge.

\end{itemize}

% What we are looking for.

\noindent{}Find:

\begin{itemize}

\item a cheapest (i.e., of the lowest cost) pair of edge-disjoint
 paths (a path is a sequence of edges), the cheaper being the working
 path, and the more expensive the protecting path,

\item continuous and contiguous units for each of the two paths
 separately: the working path and the protecting path (i.e., each
 path can have different spectrum).

\end{itemize}

% CU and incomparability.

We denote a set of contiguous units (CU) which start at index $a$ and
end at index $b$ inclusive as $\CUO{a}{b}$. For instance, $\CUO{0}{2}$
denotes units 0, 1, and 2. We can treat a set of units as a set of
CUs. For instance, $\{0, 1, 3, 4, 5\}$ and $\{\CUO{0}{1},
\CUO{3}{5}\}$ are the same. Two CUs are \emph{incomparable}, when one
is not included in the other. For instance, $\CUO{0}{2}$ and
$\CUO{2}{3}$ are incomparable, which we denote with the $\parallel$
relation, e.g., $\CUO{0}{2} \parallel \CUO{2}{3}$.

% The functions, and why to introduce them.

To state the problem generically, we intentionally introduced the
$\cost$, $\COST$, and $\decide$ functions to consider the RWA, RSA,
and RMSA problems with DPP at once. For example, for RWA, the
$\cost(p)$ function for path $p$ could return the length of the path,
for RSA, the product of the path length and the number of units
requested by the demand, and, for RMSA, the product of the path length
and the number of units required by the demand for the given path.

% The cost function for a path.

We require that the $\COST(l)$ function for a path pair $l$ be
monotonically nondecreasing, i.e., for any path pair $l'$ derived from
$l$ by appending an edge to one of the paths, \linebreak $\COST(l) \le
\COST(l')$. This requirement implies the proposed algorithm cannot be
used for networks with regeneration, when the path cost is defined as
the product of the path length and the required number of units.
Regeneration would reduce the number of required units, and the cost
of the path pair would be reduced, thus violating this requirement.

% Optimal substructure.

We also assume that an optimal path pair has the optimal substructure,
i.e., it is built of optimal path pairs, which is required by the
dynamic programming principle the proposed algorithm relies on. In
our simulations, the defined problem meets this assumption: the path
cost is the product of the path length and the number of units
required, while the cost of a path pair is the sum of the costs of the
two paths.

% On the decision function.

The $\decide$ function accepts or rejects a candidate path, and lets
the user define an acceptable path. We require the function to have
monotonically increasing requirements, i.e., if the function rejects
path $p$, then any path derived from $p$ by appending an edge should
also be rejected. For RWA, the function should make sure that the CU
has at least one unit (wavelength), for RSA, that the CU has at least
the number of units requested by the demand, and, for RMSA, that a CU
has at least the number of units required for the demand for the given
path length.

% Why is the bitrate not mentioned.

The bitrate of a demand is not a given of the stated problem and, if
needed, should be relegated to the decision function as an
implementation detail. Likewise, the $\cost$, $\COST$, and $\decide$
functions remain undefined in the problem statement. In Section
\ref{simulations}, to solve the RMSA problem with DPP, we define the
$\cost$, $\COST$, and $\decide$ functions in \linebreak Section
\ref{modulation}. The decision function defined there by
(\ref{e:decide}) checks for the required number of units, which
depends on the path length.

%%%%%%%%%%%%%%%%%%%%%%%%%%%%%%%%%%%%%%%%%%%%%%%%%%%%%%%%%%%%%%%%%%%%%%%%%%%

\section{Proposed Algorithm}
\label{algorithm}

% In a nutshell: generic Dijkstra on a search graph.

We run the generic Dijkstra algorithm on a \emph{search
graph}. Searching for a cheapest solution in the search graph
corresponds to searching for a pair of paths of lowest cost in the
input graph. The algorithm grows \emph{the search tree} for the
search graph.

\subsection{Preliminaries}

Below, we describe the search graph, the search tree, the priority
queue, and the related concepts of the solution, the path trait, and
the solution label.

\subsubsection{Search Graph}

% What a vertex in the search graph is.

The search graph has a set of vertexes $X = \{x = (v_{x, 1}, v_{x,
 2})\}$, where vertex indexes $v_{x, 1}$, and $v_{x, 2}$ of the input
graph satisfy $v_{x, 1} \le v_{x, 2}$. For vertex $x$, we find a set
of solutions, where the \emph{solution} is a pair of paths: one path
leads to vertex $v_{x, 1}$, and the other to vertex $v_{x, 2}$. Which
of the paths could eventually (when vertex $t$ is reached by both
paths) become working or protecting is unknown and unimportant at this
stage.

% What an edge in the search graph is.

An edge in the search graph from vertex $x$ to vertex $x'$ represents
finding a solution for vertex $x'$ based on a solution for vertex $x$
by taking edge $e'$ in the input graph from either vertex $v_{x, 1}$
or $v_{x, 2}$. Therefore, the edge in the search graph connects vertex
$x$ to some other vertex $x'$ which has one of the vertex indexes
$v_{x, 1}$ or $v_{x, 2}$ taken from $x$. The other vertex index of
$x'$ is the index of the target vertex of edge $e'$. Vertex $x'$
becomes either $(v_{x', 1}, v_{x, 2})$, or $(v_{x, 1}, v_{x', 2})$,
with its vertex indexes swapped if necessary, because we require the
first one be smaller than or equal to the second one.

% Appending a single edge is the simplest way.

Taking a single edge in the input graph is the simplest, and the only
one needed, way of producing a new solution in the search graph.
Taking at once two edges of the input graph, one edge for each of the
two paths, should also work, but this would lead to a more complicated and
less efficient algorithm. This is more complicated because we cannot always
take two edges, and less efficient because, by taking two edges, we can
reach a suboptimal solution, which we would avoid if we took one of
those edges first.

\subsubsection{Path Trait}

% A path trait.

A \emph{path trait} $p$ is a pair of a cost and a CU, which describes
a path in the input graph. For a path trait $p$, function $\cost(p)$
gives the path cost, and function $\CU(p)$ the path $\CU$. For
example, assuming the cost is the path length, a path trait $p = (500
\text{ km}, \CUO{0}{10})$ says the path is $\cost(p) = 500$ km long
and has the CU of $\CU(p) = \CUO{0}{10}$.

% Comparison of path traits.

Path trait $p_i$ is better than or equal to path trait $p_j$, denoted
by $p_i \le p_j$, when the cost of $p_i$ is smaller than or equal to
the cost of $p_j$, and the CU of $p_i$ includes the CU of $p_j$, i.e.,
$(\cost(p_i) \le \cost(p_j)) \land (\CU(p_i) \supseteq \CU(p_j))$. If
$p_i \le p_j$, then we drop $p_j$, since it offers no better path in
comparison with $p_i$, and so we perform the search more efficiently.

% Why we need to compare path traits in this way.

This definition of the path trait comparison allows for
\emph{incomparability} of path traits, which is needed when searching
for paths with the spectrum continuity and contiguity constraints.
For instance, path trait $p_1 = (1, \CUO{0}{2})$ is incomparable with
$p_2 = (2, \CUO{0}{3})$ because neither $p_1 \le p_2$ nor
$p_2 \le p_1$ is true. We are interested in path trait $p_2$, even
though its cost is higher than the cost of $p_1$, because $p_2$ has a
CU that is incomparable with the CU of $p_1$.

\subsubsection{Solution Label}

% What a label is.

Solution label $l_x = (p_{x, 1}, p_{x ,2})$ for vertex $x$ is a pair
of path traits $p_{x, 1}$, $p_{x ,2}$, where the first path which ends
at $v_{x, 1}$ has trait $p_{x, 1}$, and the other path which ends at
$v_{x, 2}$ has trait~$p_{x, 2}$.

% Comparison of labels.

We compare solution labels to drop those solutions which offer nothing
better than we already have, thus limiting the search space, and
performing the search more efficiently. Label $l_i$ is better than or
equal to label $l_j$, denoted by $l_i \le l_j$, when both path traits
of $l_i$ are better than or equal to both path traits of $l_j$, i.e.,
$(p_{i, 1} \le p_{j, 1}) \land (p_{i, 2} \le p_{j, 2})$. If $l_i \le
l_j$, then we should not be interested in $l_j$ because it offers no
better solution in comparison with $l_i$.

% A label which refers to a path traits leading to the same vertex.

A solution label for vertex $x$ with the same vertexes in the input
graph, i.e., \linebreak $v_{x, 1} == v_{x, 2}$, {should}
 have its path traits
ordered with the $\le$ relation, i.e., $p_{x, 1} \le p_{x, 2}$, so
that, when we compare labels of two solutions for vertex $x$, we
compare the working-path traits first, and the equal to or worse
protecting-path traits next.

% Why we need to compare solution labels in this way.

This definition of the label comparison allows for
\emph{incomparability} of labels, which is needed when searching for a
pair of paths, when these paths can have incomparable traits. For
instance, label $l_1$ of path traits $p_{1, 1} = (1, \CUO{0}{2})$, and
$p_{1, 2} = (2, \CUO{10}{12})$ is incomparable with label $l_2$ of
path traits $p_{2, 1} = (2, \CUO{0}{3})$, and
$p_{2, 2} = (10, \CUO{10}{12})$ because neither $l_1 \le l_2$ nor
$l_2 \le l_1$ is true.

\subsubsection{Search Tree}

% What a search tree is, and how we represent it.

The result of the search is the \emph{search tree}, which is organized
according to the dynamic-programming principle of reusing data from
previous computation. Search-tree node $n_{x'} = (x', l_{x'}, e',
n_x)$ represents a solution found for the search-graph vertex $x' =
(v_{x', 1}, v_{x', 2})$ based on the solution found for node $x$. The
solution is described by label $l_{x'} = (p_{x', 1}, p_{x', 2})$: the
first path of the solution which ends at $v_{x', 1}$ has trait $p_{x',
1}$, and the other path which ends at $v_{x', 2}$ has trait $p_{x',
2}$. We get the solution from the previous search-tree node $n_x$
for vertex $x$ by taking edge $e'$ in the input graph. For a
search-tree node $n$, function $\mylabel(n)$ returns its label, and
function $\vertex(n)$ returns its search-graph vertex.

% Solution types: permanent and tentative.

A tree node represents a solution which is either \emph{permanent} or
\emph{tentative}. A permanent-solution node stays in the tree for
good, while a tentative-solution node can be discarded. A
tentative-solution node is always a leaf. A tentative solution wants
to become permanent, but instead it can be discarded or never
processed.

% How we enforce edge-disjointness.

To ensure that a solution is edge-disjoint, we do not add to the
search tree a solution node if its edge was already used by its
ancestor in the search tree.

\subsubsection{Priority Queue}

% The priority queue.

The optimality of the solutions found is achieved with the priority
queue, which provides the cheapest solutions. The priority queue
stores pairs, where a pair has a cost and a reference to a search-tree
node $n_x$ of a tentative solution. The cost in the pair is the cost
of the solution, i.e., $\COST(\mylabel(n_x))$. The queue sorts
the solutions in the increasing-cost order, with the cheapest solution
at the top.

% The tentative and permanent solutions.

A tentative solution is waiting in the queue to be processed, but it
also can be either discarded, if we find a better solution, or never
processed, if the search finishes sooner. A tentative solution
becomes permanent when it is retrieved from the queue.

\subsection{Algorithm}

% The algorithm: the main loop, the relax procedure.

The proposed algorithm has the main loop listed in Algorithm
\ref{a:algorithm}, and the $\myrelax$ procedure listed in Algorithm
\ref{a:relax}. The main loop iterates over the permanent solutions
retrieved from the priority queue, while the $\myrelax$ procedure
pushes tentative solutions to the priority~queue.

% Sets of permanent and tentative labels.

The solutions for vertex $x$ are maintained in the set $P_x$ of
permanent solutions with incomparable labels, and the set $T_x$ of
tentative solutions with incomparable labels. The set of all
permanent solutions is $P$, and the set of all tentative solutions is
$T$. Permanent solutions are optimal.

% How we start the search.

We start the search at vertex $x_s = (s, s)$. We create the tentative
solution $n_{x_s}$ (the root of the search tree) of two empty paths
starting at vertex $s$ with 0 costs and the CUs of $\Omega$. We
insert $n_{x_s}$ into the set of tentative labels for vertex $x$, and
push the pair of $(0, n_{x_s})$ to the priority queue $Q$.

% Find the maximum CU, then choose the needed CU.

To cover the maximal part of the search space, we look for the paths
with the maximal CU, which satisfy the requirements of the decision
function $\decide$ used by the $\myrelax$ procedure. For this reason,
we start the search with the CU of $\Omega$.

% When the search stops.

We stop searching when the priority queue is empty, or when we find a
permanent solution for vertex $x_t = (t, t)$. If we need a complete
(i.e., for all vertexes $x$) search tree, we should let the algorithm
run until the priority queue is empty.

% The iteration of the main loop.

In each iteration of the main loop, we process the cheapest of all
tentative solutions, and make it permanent. When we retrieve a pair
from the queue, we have to ensure the tentative solution was not
discarded by the $\myrelax$ procedure, i.e., that the reference to
$n_x$ is not null.

% Relax, take two easy steps.

Next, we relax the out edges of vertex $x$ in the search graph. An
edge in the search graph represents taking an edge in the input graph
from either vertex $v_{x, 1}$ or $v_{x, 2}$, and so we iterate over
the edges leaving vertex $v_{x, 1}$ first, and over the edges leaving
vertex $v_{x, 2}$~next.

\begin{algorithm}
  \caption{Dedicated Path Protection Algorithm\\
    In: graph $G$, source vertex $s$, target vertex $t$\\
    Out: a cheapest pair of paths, and their CUs\\
    \emph{Here, we concentrate on permanent solutions $n_x$.}}
  \label{a:algorithm}
    \begin{algorithmic}
      \STATE $x_s = (s, s)$
      \STATE $x_t = (t, t)$
      \STATE $l_{x_s} = ((0, \Omega), (0, \Omega))$
      \STATE $n_{x_s} = \tn{x_s}{l_{x_s}}{\nulledge}{\text{null}}$
      \STATE $T_{x_s} = \{n_{x_s}\}$
      \STATE $\push(Q, (0, n_{x_s}))$
      \WHILE{$Q$ is not empty}
      \STATE $n_x = \pop(Q)$
      \IF{$n_x == \text{null}$}
      \STATE continue the main loop
      \ENDIF
      \STATE $x = (v_{x, 1}, v_{x, 2}) = \vertex(n_x)$
      \STATE // Remove $n_x$ from the set of tentative solutions for $x$.
      \STATE $T_x = T_x - \{n_x\}$
      \STATE // Add $n_x$ to the set of permanent solutions for $x$.
      \STATE $P_x = P_x \cup \{n_x\}$
      \IF{$x == x_t$}
      \STATE break the main loop
      \ENDIF
      \STATE $l_x = (p_{x, 1}, p_{x, 2}) = \mylabel(n_x)$
      \FORALL{out edge $e'$ of vertex $v_{x, 1}$ in $G$}
      \STATE $\myrelax(e', v_{x, 2}, p_{x, 2}, p_{x, 1}, n_x)$
      \ENDFOR
      \FORALL{out edge $e'$ of vertex $v_{x, 2}$ in $G$}
      \STATE $\myrelax(e', v_{x, 1}, p_{x, 1}, p_{x, 2}, n_x)$
      \ENDFOR
      \ENDWHILE
      \RETURN $\trace(P, x_t, x_s)$
    \end{algorithmic}
\end{algorithm}

% The relax procedure.

The $\myrelax$ procedure relaxes a single edge in the search graph,
which is described in the procedure parameters: the taken edge $e'$ in
the input graph, vertex $v_1$ and the corresponding path trait $p_1$
which both do not change, the other trait $p_2$ of the path to which
we try to append edge $e'$, and the previous search-tree node $n_x$.

% There can be a number of tentative solutions found.

The relaxation can find a number of tentative solutions, which would
differ only by the CU of $C'$, because there may be a number of
spectrum fragments available $\AU(e')$ on edge $e'$ which we can use
for a tentative solution.

% Make sure we've got the right solution.

We build (if necessary, we swap the elements of pairs $x'$ and
$l_{x'}$ with the $\swap$ function) and add a tentative solution
$n_{x'}$ to $T_{x'}$ and $Q$, only when there is no solution with a
better or equal label already found. Adding $n_{x'}$ can make some
tentative solutions invalid (since $n_{x'}$ is better), so we discard
them.

\clearpage 
\begin{algorithm}
  \caption{$\myrelax$\\
    In: edge $e'$, const vertex $v_1$, const trait $p_1$, other trait $p_2$,\\
  \hspace*{0.5 cm}previous search-tree node $n_x$\\
  \emph{Here, we concentrate on tentative solutions $n_{x'}$.}\\[-10pt]}
  \label{a:relax}
    \begin{algorithmic}
      \STATE $v' = \target(e')$
      \STATE $c' = \cost(\text{path of $p_2$ with $e'$ appended})$
      \FORALL{CU $C'$ in $\CU(p_2) \cap \AU(e')$}
      \STATE $x' = (v_1, v')$
      \STATE $p' = (c', C')$
      \IF{$\decide(p')$}
      \STATE $l_{x'} = (p_1, p')$
      \IF{$v' < v_1$}
      \STATE $\swap(x')$
      \STATE $\swap(l_{x'})$
      \ELSIF{$v_1 == v'$ and not $p_1 \le p'$}
      \STATE $\swap(l_{x'})$
      \ENDIF
      \STATE // Make sure we should be interested in $l_{x'}$.
      \IF{$\nexists n \in P_{x'} : \mylabel(n) \le l_{x'}$}
      \IF{$\nexists n \in T_{x'} : \mylabel(n) \le l_{x'}$}
      \STATE // Make sure we are not reusing $e'$.
      \IF{$e'$ not used by ancestors}
      \STATE // Discard worse tentative solutions.
      \STATE $T_{x'} = T_{x'} - \{n \in T_{x'}: l_{x'} \le \mylabel(n)\}$
      \STATE $n_{x'} = (x', l_{x'}, e', n_x)$
      \STATE // Add $n_{x'}$ to the tentative solutions for $x'$.
      \STATE $T_{x'} = T_{x'} \cup \{n_{x'}\}$
      \STATE $\push(Q, (\COST(l_{x'}), n_{x'}))$
      \ENDIF
      \ENDIF
      \ENDIF
      \ENDIF
      \ENDFOR
    \end{algorithmic}
\end{algorithm}

% Spectrum alloacation policy.

Spectrum allocation policy should be taken into account in two places.
First, the priority queue should choose the solution of the preferred
spectrum allocation policy from among the tentative solutions of the
same cost. Second, once the solution for the destination vertex is
found, the preferred CUs should be allocated from the CUs found.

% Discuss the trace function.

The $\trace$ function, using $P$, traces back the tree nodes from the
node for vertex $x_t$ to the node for vertex $x_s$. For each tree
node there is an edge, which the function appends to one of the two
paths. When appending an edge to a path, we have to ensure that not
only the cost matches, but the CU matches, too. We have to consult the
other path trait of the tree node, to ensure that we are not appending
the edge to the wrong path, which coincidentally meets the conditions.
The function returns the less expensive path as the working path, and
the more expensive path as the protecting. For each of the paths, the
function allocates the minimal CU, with the required number of units,
from the maximal CU found for the permanent solution for vertex $x_t$.

\subsection{Example}

We demonstrate how the algorithm works by finding a solution from the
source vertex $s$ to the destination vertex $t$ for a single unit in
the trap topology shown in Figure~\ref{f:ig}, where an edge label gives
not only the name of the edge, but also its cost and available units.
Not to complicate the example further, the signal modulation
constraints are not considered, especially since they are not crucial
to the algorithm as they only discard paths.

\begin{figure}[H]
  \begin{tikzpicture}

    \node [node] (s) {$s$};
    \node [node, above right = 1.5 cm and 3 cm of s] (q) {$q$};
    \node [node, below right = 1.5 cm and 3 cm of s] (r) {$r$};
    \node [node, right = 6 cm of s] (t) {$t$};

    \path (s) edge
    node [fill = white]
    {$e_1$, $\eattrs{1}{\CUO{0}{0}}$} (q);

    \path (q) edge
    node [fill = white]
    {$e_2$, $\eattrs{3}{\CUO{0}{1}}$} (t);

    \path (q) edge
    node [fill = white]
    {$e_3$, $\eattrs{1}{\CUO{0}{1}}$} (r);

    \path (s) edge
    node [fill = white]
    {$e_4$, $\eattrs{3}{\CUO{1}{1}}$} (r);

    \path (r) edge
    node [fill = white]
    {$e_5$, $\eattrs{1}{\CUO{0}{1}}$} (t);

  \end{tikzpicture}
  \caption{A sample input graph: the trap topology.}
  % Input graph.
  \label{f:ig}
\end{figure}

The trap topology is well-known since the edge-exclusion algorithm
fails for it: when the edges of the shortest path from $s$ to $t$
through vertexes $q$ and $r$ are removed, a second path does not
exist. However, the \emph{optimal solution} does exist: one path goes
through vertex $q$ with the CU of $\CUO{0}{0}$, and the other through
vertex $r$ with $\CUO{1}{1}$.

Figure~\ref{f:sg} shows the search graph generated, where the edge label
gives the name of an input-graph edge to take to make a transition
between the vertexes in the search graph. For example, the transition
from vertex $(s, s)$ to vertex $(q, s)$ requires edge $e_1$. Most
edges are undirected, since the algorithm examines transitions in both
directions. However, there are some directed edges (e.g., from $(q,
s)$ to $(s, t)$), since their reverse transitions (e.g., from $(s, t)$
to $(q, s)$) are not examined. Because paths in the input graph have
to be edge-disjoint, some paths in the search graph are disallowed,
e.g., $(s, s) - (q, s) - (q, q)$.

% start a new page without indent 4.6cm
%\clearpage
\end{paracol}
\nointerlineskip

\newlength{\gd}
\setlength{\gd}{2.5 cm}

\begin{figure}[H]\widefigure
  \begin{tikzpicture}[> = stealth]
    
    % Nodes ---------------------------------------------------

    \begin{scope}[every node/.style = {draw, circle}]
      
      \node (vss) at (0, 0) {$(s, s)$};

      \node (vqs) at (\gd, \gd) {$(q, s)$};
      \node (vrs) at (\gd, -\gd) {$(r, s)$};

      \node (vqr) at (2\gd, 0) {$(q, r)$};

      \node (vqq) at (3\gd, \gd) {$(q, q)$};
      \node (vrr) at (3\gd, -\gd) {$(r, r)$};

      \node (vst) at (4\gd, 0) {$(s, t)$};

      \node (vqt) at (5\gd, \gd) {$(q, t)$};
      \node (vrt) at (5\gd, -\gd) {$(r, t)$};

      \node (vtt) at (6\gd, 0) {$(t, t)$};

    \end{scope}

    % Edges ---------------------------------------------------

    \draw (vss) to node [circle, fill = white] {$e_1$} (vqs);
    \draw (vss) to node [circle, fill = white] {$e_4$} (vrs);
    \draw (vqs) to node [circle, fill = white] {$e_3$} (vrs);

    \draw (vqs) to node [circle, fill = white] {$e_1$} (vqq);
    \draw (vqs) to node [circle, fill = white] {$e_4$} (vqr);
    \draw [->] (vqs) to node [circle, fill = white, near start] {$e_2$} (vst);
    \draw [->] (vrs) to node [circle, fill = white, near start] {$e_5$} (vst);
    \draw (vrs) to node [circle, fill = white] {$e_1$} (vqr);
    \draw (vrs) to node [circle, fill = white] {$e_4$} (vrr);

    \draw [->] (vqq) to node [circle, fill = white] {$e_2$} (vqt);
    \draw (vqr) to node [circle, fill = white, near end] {$e_3$} (vqq);
    \draw [->] (vqr) to node [circle, fill = white] {$e_5$} (vqt);
    \draw (vst) to node [circle, fill = white] {$e_1$} (vqt);
    \draw (vst) to node [circle, fill = white] {$e_4$} (vrt);
    \draw [->] (vqr) to node [circle, fill = white] {$e_2$} (vrt);
    \draw (vqr) to node [circle, fill = white, near end] {$e_3$} (vrr);
    \draw [->] (vrr) to node [circle, fill = white] {$e_5$} (vrt);

    \draw[->] (vqt) to node [circle, fill = white] {$e_2$} (vtt);
    \draw[->] (vrt) to node [circle, fill = white] {$e_5$} (vtt);
    \draw (vqt) to node [circle, fill = white] {$e_3$} (vrt);
    
  \end{tikzpicture}
  \caption{The search graph.}
  % Search graph.
  \label{f:sg}
\end{figure}
\begin{paracol}{2}
%\linenumbers
\switchcolumn

Figure~\ref{f:st} shows the search tree generated, where only the
permanent (and not tentative) solutions are represented. The tree is
rooted at node $n_1$ for vertex $(s, s)$. The solution found is
represented by node $n_{14}$ for vertex $(t, t)$. A search-graph
vertex can have a set of permanent incomparable solutions, which are
represented by a set of search-tree nodes, e.g., for vertex $(q, s)$
there are two search-tree nodes: $n_2$ and $n_8$.

The algorithm processes solution labels by taking actions on them, as
reported in Table \ref{t:sl}. A label can be produced and pushed into
the queue, as in, e.g., action \#0. A row represents an action on a
label which was produced for the given search-graph vertex by making a
transition with the given edge. For instance, action \#2 reports a
label that is pushed into the queue, and which was produced for the
search-graph vertex $(q, s)$ by making a transition with edge $e_1$
(from vertex $(s, s)$).

A label can be retrieved from the queue, and made permanent, as in,
e.g., action \#1. A search-tree node for a permanent label has a name
reported, e.g., $n_1$ as in action \#1. A row for a label made
permanent is colored gray to mark the beginning of a sequence of rows
that report the actions of the relaxation based on the label from that
gray row. For instance, the row for action \#1 is gray, and the
subsequent rows for actions from \#2 to \#5 report the actions of the
relaxation that ensued.

% start a new page without indent 4.6cm
%\clearpage
\end{paracol}
\nointerlineskip

\begin{figure}[H]\widefigure
  \begin{tikzpicture}[> = stealth, els/.style = {circle, fill = white}]
    
    % Nodes ---------------------------------------------------

    % Distance above a node to a label.
    \newlength{\dan}
    \setlength{\dan}{0.0 cm}

    \begin{scope}[vsn/.style = {draw, rectangle split,
          rectangle split parts = #1, anchor = center,
          minimum width = 1 cm}]
      
      \node[vsn = 1] (vss) at (0, 0) {$n_1$};
      \node[above = \dan of vss.north] {$(s, s)$};

      \node[vsn = 2] (vqs) at (\gd, \gd) {$n_2$\nodepart{two}$n_8$};
      \node[above = \dan of vqs.north] {$(q, s)$};
      \node[vsn = 2] (vrs) at (\gd, -\gd) {$n_3$\nodepart{two}$n_5$};
      \node[above = \dan of vrs.north] {$(r, s)$};

      \node[vsn = 1] (vqr) at (2\gd, 0) {$n_7$};
      \node[above = \dan of vqr.north] {$(q, r)$};

      \node[vsn = 1] (vqq) at (3\gd, \gd) {$n_{10}$};
      \node[above = \dan of vqq.north] {$(q, q)$};
      \node[vsn = 1] (vrr) at (3\gd, -\gd) {$n_{11}$};
      \node[above = \dan of vrr.north] {$(r, r)$};

      \node[vsn = 2] (vst) at (4\gd, 0) {$n_4$\nodepart{two}$n_6$};
      \node[above = \dan of vst.north] {$(s, t)$};

      \node[vsn = 1] (vqt) at (5\gd, \gd) {$n_9$};
      \node[above = \dan of vqt.north] {$(q, t)$};
      \node[vsn = 2] (vrt) at (5\gd, -\gd) {$n_{12}$\nodepart{two}$n_{13}$};
      \node[above = \dan of vrt.north] {$(r, t)$};

      \node[vsn = 1] (vtt) at (6\gd, 0) {$n_{14}$};
      \node[above = \dan of vtt.north] {$(t, t)$};

    \end{scope}

    % Edges ---------------------------------------------------
    
    // From (s, s)
    \draw[->] (vss.text east) to [out = 0, in = 180] node [els] {$e_1$} (vqs.text west);
    \draw[->] (vss.text east) to [out = 0, in = 180] node [els] {$e_4$} (vrs.two west);

    // From (q, s)
    \draw[->] (vqs.text east) to [out = 0, in = 180] node [els] {$e_4$} (vqr.text west);
    \draw[->] (vqs.text east) to [out = 0, in = 180] node [els, pos = 0.65] {$e_3$} (vrs.text west);

    // From (r, s)
    \draw[->] (vrs.text east) to [out = 0, in = 180] node [els] {$e_5$} (vst.text west);
    \draw[->] (vrs.text east) to [out = 0, in = 180] node [els, near end] {$e_4$} (vrr.text west);
    \draw[->] (vrs.two east) to [out = 0, in = 180] node [els, pos = 0.65] {$e_3$} (vqs.two west);
    \draw[->] (vrs.two east) to [out = 0, in = 180] node [els] {$e_5$} (vst.two west);

    // From (q, r)
    \draw[->] (vqr.text east) to [out = 0, in = 180] node [els] {$e_3$} (vqq.text west);

    // From (s, t)
    \draw[->] (vst.text east) to [out = 0, in = 180] node [els] {$e_4$} (vrt.two west);
    \draw[->] (vst.two east) to [out = 0, in = 180] node [els] {$e_1$} (vqt.two west);

    // From (q, t)
    \draw[->] (vqt.text east) to [out = 0, in = 180] node [els] {$e_2$} (vtt.text west);
    \draw[->] (vqt.two east) to [out = 0, in = 180] node [els] {$e_3$} (vrt.text west);

  \end{tikzpicture}
  \caption{The search tree.}
  % Search tree.
  \label{f:st}
\end{figure}
\begin{paracol}{2}
%\linenumbers
\switchcolumn

A label can be pushed into the queue, or it can be dropped for two
reasons: it is worse than or equal to an existing label (e.g., in
action \#4, a label is dropped because an equal label of action \#2
exists), or it uses an edge twice (e.g., in action \#7, a label is
dropped because edge $e_1$ is used twice). A label can also be
discarded from the queue if a better label is found (e.g., in action
\#17, a label is discarded because a better label was found in action
\#16).

As reported in Table \ref{t:sl}, the search is booted with action \#0.
There are 14 permanent labels found for 10 vertexes of the search
graph. The algorithm terminates, when the destination node $(t, t)$
is reached with the search-tree node $n_{14}$. We can trace back from
$n_{14}$ to $n_1$ to get the aforementioned optimal solution.

% start a new page without indent 4.6cm

\end{paracol}
\nointerlineskip

\begin{specialtable}[H]
\widetable  \setlength{\tabcolsep}{3.75mm}
\caption{Solution labels processed.}
  % Solution labels.
  \label{t:sl}
  \small
  \begin{tabular}{ccccccl}
    \toprule
    \begin{tabular}{@{}c@{}}\textbf{Action}\\\textbf{Number}\end{tabular} &
    \begin{tabular}{@{}c@{}}\textbf{Solution}\\\textbf{Cost}\end{tabular} &
    \begin{tabular}{@{}c@{}}\textbf{Search-Tree}\\\textbf{Node Name}\end{tabular} &
    \begin{tabular}{@{}c@{}}\textbf{Search-Graph}\\\textbf{Vertex}\end{tabular} &
    \textbf{Solution Label }& \textbf{Edge }&\textbf{ Action}\\
    \hline
    0 & 0 &         & $(s, s)$ & $((0, \CUO{0}{1}), (0, \CUO{0}{1}))$ & $e_{\emptyset}$ & push into queue\\
    \rowcolor{lightgray}\hline
    1 & 0 & $n_1$   & $(s, s)$ & $((0, \CUO{0}{1}), (0, \CUO{0}{1}))$ & $e_{\emptyset}$ & make permanent\\
    2 & 1 &         & $(q, s)$ & $((1, \CUO{0}{0}), (0, \CUO{0}{1}))$ & $e_1$ & push into queue\\
    3 & 3 &         & $(r, s)$ & $((3, \CUO{1}{1}), (0, \CUO{0}{1}))$ & $e_4$ & push into queue\\
    4 & 1 &         & $(q, s)$ & $((1, \CUO{0}{0}), (0, \CUO{0}{1}))$ & $e_1$ & drop (worse or equal)\\
    5 & 3 &         & $(r, s)$ & $((3, \CUO{1}{1}), (0, \CUO{0}{1}))$ & $e_4$ & drop (worse or equal)\\
    \rowcolor{lightgray}\hline
    6 & 1 & $n_2$   & $(q, s)$ & $((1, \CUO{0}{0}), (0, \CUO{0}{1}))$ & $e_1$ & make permanent\\
    7 & 2 &         & $(q, q)$ & $((1, \CUO{0}{0}), (1, \CUO{0}{0}))$ & $e_1$ & drop (edge reuse)\\
    8 & 4 &         & $(q, r)$ & $((1, \CUO{0}{0}), (3, \CUO{1}{1}))$ & $e_4$ & push into queue\\
    9 & 2 &         & $(s, s)$ & $((0, \CUO{0}{1}), (2, \CUO{0}{0}))$ & $e_1$ & drop (worse or equal)\\
   10 & 4 &         & $(s, t)$ & $((0, \CUO{0}{1}), (4, \CUO{0}{0}))$ & $e_2$ & push into queue\\
   11 & 2 &         & $(r, s)$ & $((2, \CUO{0}{0}), (0, \CUO{0}{1}))$ & $e_3$ & push into queue\\
    \rowcolor{lightgray}\hline
   12 & 2 & $n_3$   & $(r, s)$ & $((2, \CUO{0}{0}), (0, \CUO{0}{1}))$ & $e_3$ & make permanent\\
   13 & 3 &         & $(q, r)$ & $((1, \CUO{0}{0}), (2, \CUO{0}{0}))$ & $e_1$ & drop (edge reuse)\\
   14 & 5 &         & $(r, r)$ & $((2, \CUO{0}{0}), (3, \CUO{1}{1}))$ & $e_4$ & push into queue\\
   15 & 3 &         & $(q, s)$ & $((3, \CUO{0}{0}), (0, \CUO{0}{1}))$ & $e_3$ & drop (worse or equal)\\
   16 & 3 &         & $(s, t)$ & $((0, \CUO{0}{1}), (3, \CUO{0}{0}))$ & $e_5$ & push into queue\\
   17 & 4 &         & $(s, t)$ & $((0, \CUO{0}{1}), (4, \CUO{0}{0}))$ & $e_2$ & discard from queue\\
   \bottomrule
  \end{tabular}
\end{specialtable}

\begin{specialtable}[H]\ContinuedFloat
\widetable\small  \setlength{\tabcolsep}{3.75mm}
\caption{{\em Cont.}}
 \begin{tabular}{ccccccl}
    \toprule
    \begin{tabular}{@{}c@{}}\textbf{Action}\\\textbf{Number}\end{tabular} &
    \begin{tabular}{@{}c@{}}\textbf{Solution}\\\textbf{Cost}\end{tabular} &
    \begin{tabular}{@{}c@{}}\textbf{Search-Tree}\\\textbf{Node Name}\end{tabular} &
    \begin{tabular}{@{}c@{}}\textbf{Search-Graph}\\\textbf{Vertex}\end{tabular} &
    \textbf{Solution Label }& \textbf{Edge }&\textbf{ Action}\\
      \rowcolor{lightgray}\hline
   18 & 3 & $n_4$   & $(s, t)$ & $((0, \CUO{0}{1}), (3, \CUO{0}{0}))$ & $e_5$ & make permanent\\
   19 & 4 &         & $(q, t)$ & $((1, \CUO{0}{0}), (3, \CUO{0}{0}))$ & $e_1$ & drop (edge reuse)\\
   20 & 6 &         & $(r, t)$ & $((3, \CUO{1}{1}), (3, \CUO{0}{0}))$ & $e_4$ & push into queue\\
    \rowcolor{lightgray}\hline
   21 & 3 & $n_5$   & $(r, s)$ & $((3, \CUO{1}{1}), (0, \CUO{0}{1}))$ & $e_4$ & make permanent\\
   22 & 4 &         & $(q, r)$ & $((1, \CUO{0}{0}), (3, \CUO{1}{1}))$ & $e_1$ & drop (worse or equal)\\
   23 & 6 &         & $(r, r)$ & $((3, \CUO{1}{1}), (3, \CUO{1}{1}))$ & $e_4$ & drop (edge reuse)\\
   24 & 4 &         & $(q, s)$ & $((4, \CUO{1}{1}), (0, \CUO{0}{1}))$ & $e_3$ & push into queue\\
   25 & 6 &         & $(s, s)$ & $((0, \CUO{0}{1}), (6, \CUO{1}{1}))$ & $e_4$ & drop (worse or equal)\\
   26 & 4 &         & $(s, t)$ & $((0, \CUO{0}{1}), (4, \CUO{1}{1}))$ & $e_5$ & push into queue\\
    \rowcolor{lightgray}\hline
   27 & 4 & $n_6$   & $(s, t)$ & $((0, \CUO{0}{1}), (4, \CUO{1}{1}))$ & $e_5$ & make permanent\\
   28 & 5 &         & $(q, t)$ & $((1, \CUO{0}{0}), (4, \CUO{1}{1}))$ & $e_1$ & push into queue\\
   29 & 7 &         & $(r, t)$ & $((3, \CUO{1}{1}), (4, \CUO{1}{1}))$ & $e_4$ & drop (edge reuse)\\
    \rowcolor{lightgray}\hline
   30 & 4 & $n_7$   & $(q, r)$ & $((1, \CUO{0}{0}), (3, \CUO{1}{1}))$ & $e_4$ & make permanent\\
   31 & 5 &         & $(r, s)$ & $((3, \CUO{1}{1}), (2, \CUO{0}{0}))$ & $e_1$ & drop (worse or equal)\\
   32 & 7 &         & $(r, t)$ & $((3, \CUO{1}{1}), (4, \CUO{0}{0}))$ & $e_2$ & drop (worse or equal)\\
   33 & 5 &         & $(r, r)$ & $((2, \CUO{0}{0}), (3, \CUO{1}{1}))$ & $e_3$ & drop (worse or equal)\\
   34 & 5 &         & $(q, q)$ & $((1, \CUO{0}{0}), (4, \CUO{1}{1}))$ & $e_3$ & push into queue\\
   35 & 7 &         & $(q, s)$ & $((1, \CUO{0}{0}), (6, \CUO{1}{1}))$ & $e_4$ & drop (worse or equal)\\
   36 & 5 &         & $(q, t)$ & $((1, \CUO{0}{0}), (4, \CUO{1}{1}))$ & $e_5$ & drop (worse or equal)\\
    \rowcolor{lightgray}\hline
   37 & 4 & $n_8$   & $(q, s)$ & $((4, \CUO{1}{1}), (0, \CUO{0}{1}))$ & $e_3$ & make permanent\\
   38 & 5 &         & $(q, q)$ & $((1, \CUO{0}{0}), (4, \CUO{1}{1}))$ & $e_1$ & drop (worse or equal)\\
   39 & 7 &         & $(q, r)$ & $((4, \CUO{1}{1}), (3, \CUO{1}{1}))$ & $e_4$ & drop (edge reuse)\\
   40 & 7 &         & $(s, t)$ & $((0, \CUO{0}{1}), (7, \CUO{1}{1}))$ & $e_2$ & drop (worse or equal)\\
   41 & 5 &         & $(r, s)$ & $((5, \CUO{1}{1}), (0, \CUO{0}{1}))$ & $e_3$ & drop (worse or equal)\\
    \rowcolor{lightgray}\hline
   42 & 5 & $n_9$   & $(q, t)$ & $((1, \CUO{0}{0}), (4, \CUO{1}{1}))$ & $e_1$ & make permanent\\
   43 & 6 &         & $(s, t)$ & $((2, \CUO{0}{0}), (4, \CUO{1}{1}))$ & $e_1$ & drop (worse or equal)\\
   44 & 8 &         & $(t, t)$ & $((4, \CUO{0}{0}), (4, \CUO{1}{1}))$ & $e_2$ & push into queue\\
   45 & 6 &         & $(r, t)$ & $((2, \CUO{0}{0}), (4, \CUO{1}{1}))$ & $e_3$ & push into queue\\
    \rowcolor{lightgray}\hline
   46 & 5 & $n_{10}$ & $(q, q)$ & $((1, \CUO{0}{0}), (4, \CUO{1}{1}))$ & $e_3$ & make permanent\\
   47 & 6 &         & $(q, s)$ & $((4, \CUO{1}{1}), (2, \CUO{0}{0}))$ & $e_1$ & drop (worse or equal)\\
   48 & 8 &         & $(q, t)$ & $((4, \CUO{1}{1}), (4, \CUO{0}{0}))$ & $e_2$ & push into queue\\
   49 & 6 &         & $(q, r)$ & $((4, \CUO{1}{1}), (2, \CUO{0}{0}))$ & $e_3$ & drop (edge reuse)\\
   50 & 8 &         & $(q, t)$ & $((1, \CUO{0}{0}), (7, \CUO{1}{1}))$ & $e_2$ & drop (worse or equal)\\
   51 & 6 &         & $(q, r)$ & $((1, \CUO{0}{0}), (5, \CUO{1}{1}))$ & $e_3$ & drop (worse or equal)\\
    \rowcolor{lightgray}\hline
   52 & 5 & $n_{11}$ & $(r, r)$ & $((2, \CUO{0}{0}), (3, \CUO{1}{1}))$ & $e_4$ & make permanent\\
   53 & 6 &         & $(q, r)$ & $((3, \CUO{0}{0}), (3, \CUO{1}{1}))$ & $e_3$ & drop (worse or equal)\\
   54 & 6 &         & $(r, t)$ & $((3, \CUO{1}{1}), (3, \CUO{0}{0}))$ & $e_5$ & drop (worse or equal)\\
   55 & 6 &         & $(q, r)$ & $((4, \CUO{1}{1}), (2, \CUO{0}{0}))$ & $e_3$ & drop (edge reuse)\\
   56 & 8 &         & $(r, s)$ & $((2, \CUO{0}{0}), (6, \CUO{1}{1}))$ & $e_4$ & drop (worse or equal)\\
   57 & 6 &         & $(r, t)$ & $((2, \CUO{0}{0}), (4, \CUO{1}{1}))$ & $e_5$ & drop (worse or equal)\\
    \rowcolor{lightgray}\hline
   58 & 6 & $n_{12}$ & $(r, t)$ & $((2, \CUO{0}{0}), (4, \CUO{1}{1}))$ & $e_3$ & make permanent\\
   59 & 7 &         & $(q, t)$ & $((3, \CUO{0}{0}), (4, \CUO{1}{1}))$ & $e_3$ & drop (worse or equal)\\
   60 & 7 &         & $(t, t)$ & $((3, \CUO{0}{0}), (4, \CUO{1}{1}))$ & $e_5$ & drop (edge reuse)\\
    \rowcolor{lightgray}\hline
   61 & 6 & $n_{13}$ & $(r, t)$ & $((3, \CUO{1}{1}), (3, \CUO{0}{0}))$ & $e_4$ & make permanent\\
   62 & 7 &         & $(q, t)$ & $((4, \CUO{1}{1}), (3, \CUO{0}{0}))$ & $e_3$ & drop (edge reuse)\\
   63 & 9 &         & $(s, t)$ & $((6, \CUO{1}{1}), (3, \CUO{0}{0}))$ & $e_4$ & drop (worse or equal)\\
   64 & 7 &         & $(t, t)$ & $((3, \CUO{0}{0}), (4, \CUO{1}{1}))$ & $e_5$ & drop (edge reuse)\\
    \rowcolor{lightgray}\hline
   65 & 8 & $n_{14}$ & $(t, t)$ & $((4, \CUO{0}{0}), (4, \CUO{1}{1}))$ & $e_2$ & make permanent\\
    \hline
  \end{tabular}

\end{specialtable}

\begin{paracol}{2}
%\linenumbers
\switchcolumn

\clearpage 

\subsection{Worst-Case Analysis}

% Search space is limited.

We argue the size of the search space is polynomially upper bounded.
We derive the upper bound $L$ of the number of incomparable labels
(i.e., the size of the search space) by considering the worst case
where every vertex of all $|X|$ search graph vertexes has the maximum
number $S$ of incomparable labels. Therefore, $L = |X|S$. The problem
is to derive $|X|$ and $S$.

% The number of vertexes.

The number of vertexes in the search graph is given by
(\ref{e:nvertexes}), since the input graph has $|V|$ vertexes, and
since vertexes $x = (v_{x, 1}, v_{x, 2})$ of the search graph satisfy
$v_{x, 1} \le v_{x, 2}$. The number of vertexes with $v_{x, 1} =
v_{x, 2}$ is $|V|$, and the number of vertexes with $v_{x, 1} < v_{x,
 2}$ is the number of combinations of two elements from the set of
$|V|$ elements.

\begin{equation}
  |X| = |V| + \binom{|V|}{2} = \frac{|V|(|V| + 1)}{2}.
  \label{e:nvertexes}
\end{equation}

% The number of incomparable labels.

The maximum number $S$ of incomparable labels a vertex can have is
given by (\ref{e:nlabels}). Since a label describes a solution made
up of two independent paths, $S$ is the maximum number of incomparable
path traits squared.

% The number of incomparable traits.

The maximum number of incomparable path traits depends only on
$\Omega$. We get the largest set of incomparable path traits when the
cost of path traits increases as the size of their CUs increases. The
largest set has $|\Omega|$ subsets: the first subset has $|\Omega|$
traits with CUs of a single unit and the lowest cost, the second
subset has $|\Omega| - 1$ traits with CUs of two units and a higher
cost, \ldots, and the last subset has a single trait with the CU of
$|\Omega|$ units and the highest cost. The largest set has $1 + 2 +
\ldots + |\Omega| = (|\Omega| + 1)|\Omega|/2$ incomparable path
traits.

\begin{equation}
  S = \left(\frac{(|\Omega| + 1)|\Omega|}{2}\right)^2.
  \label{e:nlabels}
\end{equation}

Therefore, the size of the search space is polynomially upper bounded,
since \linebreak $O(L) = O(|V|^2|\Omega|^4)$.

%%%%%%%%%%%%%%%%%%%%%%%%%%%%%%%%%%%%%%%%%%%%%%%%%%%%%%%%%%%%%%%%%%%%%%%%%%%

\section{Simulations}
\label{simulations}

The simulations had two goals: the optimality corroboration, and the
performance evaluation. We had 48,600 simulation runs: 32,400
corroborative runs, and 16,200 performance evaluation runs.

We corroborated the optimality of the results of our algorithm by
comparing them with the results of the brute-force enumeration
algorithm: for a single search either both algorithms returned results
of the same cost, or both algorithms returned no results. Since there
are billions of feasible solutions even in small networks, and since
the brute-force algorithm enumerates them all, we were able to
corroborate the results only for small~networks.

\textls[-15]{We ran the corroborative simulations for networks of 10, 11, 12, 13,
14, and 15 nodes; for 160, 320, and 640 units; for offered loads
ranging from light to heavy; and for demands requesting on average
from 10 to 64 units. In total, we had 32,400 simulation runs, out of
which 183 runs were killed, because they requested more than 120 GB of
operating memory, which we did not have. In total, we carried out
17,590,624 searches, all successfully~corroborated.}

The remainder of this section is about the performance evaluation.

\subsection{Simulation Setting}
\label{setting}

Below, we describe how we model the network, the traffic, and the
signal modulation.

\subsubsection{Network Model}

A network model has an undirected graph, and $|\Omega|$. We randomly
generated three groups of network graphs with 25, 50, and 100
vertexes, where each group had one hundred graphs. We generated
Gabriel graphs because they have been shown to model the properties
of the transport networks very well \cite{10.1109/ICUMT.2013.6798402}.
The vertexes were uniformly distributed over a square area with the
density of one vertex per 10 thousand square km.

% Spacings.

We used three spacings of 25 GHz, 12.5 GHz, and 6.25 GHz for the
erbium band, which translated to three values for $|\Omega|$: 160,
320, and 640 units.

% Spectrum allocation policy.

We present the results only for the first-fit spectrum allocation
policy. The first-fit policy allocates units in the first CU that can
support the demand, i.e., the CU with the units of the lowest indexes.
We also considered the best-fit and random-fit policies. The best-fit
policy performed comparably to the first-fit policy, and the
random-fit policy performed markedly worse than the first-fit policy.
We do not present the results for these alternative policies because
they add little to the main results.

\subsubsection{Traffic Model}

% The network utilization, and the offered load.

We evaluate the algorithm performance as a function of the
\emph{network utilization}, which we define as the ratio of the number
of units in use to the total number of units on all edges. We measure
the network utilization in response to offered load $a$, which
expresses the desired network utilization.

% Demands characteristics.

Demands arrive according to the exponential distribution with rate
$\lambda$ per day. The end nodes of a demand are different and chosen
at random. The number of units a demand requests is described by
distribution $(\Poisson(\gamma - 1) + 1)$ with the mean of $\gamma$,
i.e., a shifted Poisson distribution, so that we do not get a zero.
Parameter $\gamma_p$ expresses the mean number of units that demands
request relative to the number $|\Omega|$ of all units on every edge,
and so $\gamma = \gamma_p |\Omega|$. We model the connection holding
time with the exponential distribution with the mean of $\tau$ days.
A connection is bidirectional: the same CU is allocated in both
directions for a path.

% Lambda as a function of the offered load.

We express $\lambda$ as a function of $a$. The offered load is the
ratio of the number of demanded units to the total number of units on
all edges. For traffic intensity $\lambda\tau$, the number of units
demanded is $2\lambda\tau\gamma\alpha$, since a demand requests two
paths, and we estimate they require $\gamma$ units, and $\alpha$ edges
each, where $\alpha$ is the average number of edges of a shortest path
between the end nodes of the demand in the network being simulated.
Therefore, $a = 2\lambda\tau\gamma\alpha / |E||\Omega|$, from which
(\ref{e:lambda}) follows.

\begin{equation}
  \lambda(a) = \frac{a|E||\Omega|}{2\tau\gamma\alpha}.
  \label{e:lambda}
\end{equation}

Equation (\ref{e:lambda}) underestimates the value of $\lambda(a)$
because we assume that every demand has a connection established. For
this reason, $a = 1$ does not yield a full network utilization.

\subsubsection{Signal Modulation Model}
\label{modulation}

% The signal modulation in the EON.

We use the signal modulation model from Reference \cite{10.1364/JOCN.11.000568},
with $M$ modulations available. For a demand requesting $g$ units for
the most spectrally-efficient modulation, the number of units needed
to establish a connection of length $d$ is given by (\ref{e:u}), where
$r_1$ is the reach of the least spectrally-efficient modulation, and
$r_M$ is the reach of the most spectrally-efficient modulation.

\begin{equation}
  u(g, d) =
  \begin{cases}
    g & \text{if}\ d \le r_M \\
    \infty & \text{if}\ r_1 < d \\
    \ceil{g \cdot log_2(2d / r_M)} & \text{otherwise}%
  \end{cases}.
  \label{e:u}
\end{equation}

% We could use bitrates too, instead of the number of units.

We describe a demand with the number of units $g$, instead of bitrate
$b$, because the algorithm works with units, not bitrates. If the
bitrate is given, we can calculate the number of units using
(\ref{e:g}), where $R$ is a technology-dependent bitrate (e.g., $2.5$
Gb/s), and $G$ is the number of guard-band units
\cite{10.1364/JOCN.4.000603}.

\begin{equation}
  g(b) = \ceil{b/(R \cdot M)} + G.
  \label{e:g}
\end{equation}

% On $r_M$.

In the simulations, we assumed $M = 4$, and the reach of the
least-spectrally efficient modulation $r_1$ equals to one and a half
lengths of the longest path from among all the shortest paths (i.e.,
for every source-destination combination) in the network being
simulated, which allows us to consider paths much longer than an
average shortest path. \linebreak Following Reference \cite{10.1364/JOCN.11.000568}, we
calculated $r_M = r_1 / 2^{M - 1}$.

\subsubsection{The Cost and Decision Functions}

% The cost and decision functions.

The cost and decision functions for path $p$ are given by
(\ref{e:cost}) and (\ref{e:decide}), where function $\length$ returns
the length of path $p$ as the sum of positive lengths of the edges
used. The cost function for path pair $l$ is given by (\ref{e:COST}).

\begin{equation}
  \cost(p) = \length(p) \cdot u(g, \length(p)),
  \label{e:cost}
\end{equation}
\begin{equation}
  \decide(p) = u(g, \length(p)) \le |\CU(p)|,
  \label{e:decide}
\end{equation}
\begin{equation}
  \COST(l = (p_1, p_2)) = \cost(p_1) + \cost(p_2).
  \label{e:COST}
\end{equation}

\subsection{Runs and Populations}

% What a simulation run is.

\emph{A simulation run} simulated 150 days of a network in operation,
with the results from the first 50 days discarded. The parameters of
a simulation run were: the network size, $|\Omega|$, $\gamma$, $a$,
and $\tau$. A simulation run reported the mean network utilization,
the mean and maximum times taken, and the mean and maximum number of
64-bit memory words used by a search for a single demand.

% Simulation results -> sample results -> population results

We averaged the mean simulation results to calculate the \emph{sample}
mean results, which estimate the \emph{population} mean results, and
the average algorithm performance. We took the maximum of the maximum
simulation results to get the \emph{sample} maximum results, which
estimate the population maximum results, and the worst-case algorithm
performance.

% The parameters of a population, and the details.

In a given population, there were 100 simulation runs whose parameters
differed only with the network model. We had 162 populations because
we varied 3 network sizes (25, 50, 100 nodes), 3 values of $|\Omega|$
(160, 320, 640 units), 9 values of $a$ (0.05, 0.1, 0.15, 0.2, 0.45,
0.65, 1, 1.5, 2), and two runs for $\gamma = 10$ units, and $\gamma_p
= 10\%$ of units available (i.e., 16 units for the case with 160
units, 32 for 320, and 64 for 640). For all populations, the mean
connection holding time $\tau = 10$ days was constant. In total, we
carried out 16200 simulation runs (162 populations $\times$ 100
samples) with 24043157 searches. The sample means credibly estimate
the population means, since their relative standard error was below
5\%.

%%%%%%%%%%%%%%%%%%%%%%%%%%%%%%%%%%%%%%%%%%%%%%%%%%%%%%%%%%%%%%%%%%%%%%%%%%%

\subsection{Simulation Results}
\label{results}

Figures \ref{f:time} and \ref{f:memory} show the sample means and the
sample maxima of the time taken and memory used by a search,
regardless of whether the search was successful or not. The results
are shown on a logarithmic scale as a function of network utilization.
The curves are plotted dotted for 160 units, dashed for 320 units, and
solid for 640 units. The sample means are plotted thin, and the
sample maxima thick. Each curve is drawn using 9 data points for
different values of $a$. For the means, we do not report the error
bars representing the standard error, since they were too small to
plot.

% The subfigures.

Figures \ref{f:time} and \ref{f:memory} have three rows and two
columns of subfigures. The first row shows the results for the
networks with 25 nodes, the second for 50 nodes, and the third for 100~nodes. The first column shows the results for $\gamma = 10$, and the
second column the results for $\gamma_p = 10 \%$.

% The time

% BBP: discussion.
% start a new page without indent 4.6cm
\clearpage
\end{paracol}
\nointerlineskip
\begin{figure}[H]\widefigure
  \subfloat[time taken for $\gamma = 10$, and 25 nodes]{%
    \input{Definitions/time_10_25}}\hfill%
  \subfloat[time taken for $\gamma_p = 10\%$, and 25 nodes]{%
    \input{Definitions/time_10p_25}}\\[5pt]%
  \subfloat[time taken for $\gamma = 10$, and 50 nodes]{%
    \input{Definitions/time_10_50}}\hfill%
  \subfloat[time taken for $\gamma_p = 10\%$, and 50 nodes]{%
    \input{Definitions/time_10p_50}}\\[5pt]%
  \subfloat[time taken for $\gamma = 10$, and 100 nodes]{%
    \input{Definitions/time_10_100}}\hfill%
  \subfloat[time taken for $\gamma_p = 10\%$, and 100 nodes]{%
    \input{Definitions/time_10p_100}}\\[5pt]%
  \centering\ref{regular}%
  \caption{The sample means and maxima of the time taken by the
    proposed algorithm.}
  \label{f:time}
\end{figure}
\begin{paracol}{2}
%\linenumbers
\switchcolumn

% start a new page without indent 4.6cm
%\clearpage
\end{paracol}
\nointerlineskip
\begin{figure}[H]\widefigure
  \subfloat[memory used for $\gamma = 10$, and 25 nodes]{%
    \input{Definitions/memory_10_25}}\hfill%
  \subfloat[memory used for $\gamma_p = 10\%$, and 25 nodes]{%
    \input{Definitions/memory_10p_25}}\\[5pt]%
  \subfloat[memory used for $\gamma = 10$, and 50 nodes]{%
    \input{Definitions/memory_10_50}}\hfill%
  \subfloat[memory used for $\gamma_p = 10\%$, and 50 nodes]{%
    \input{Definitions/memory_10p_50}}\\[5pt]%
  \subfloat[memory used for $\gamma = 10$, and 100 nodes]{%
    \input{Definitions/memory_10_100}}\hfill%
  \subfloat[memory used for $\gamma_p = 10\%$, and 100 nodes]{%
    \input{Definitions/memory_10p_100}}\\[5pt]%
  \centering\ref{regular}%
  \caption{The sample means and maxima of the memory used by the
    proposed algorithm.}
  \label{f:memory}
\end{figure}
\begin{paracol}{2}
%\linenumbers
\switchcolumn

The mean times range from $10^{-3}$ s (for 25 nodes, and 160 units) to
$10^2$ s (100 nodes, 640 units). While the difference in scale is
$10^5$, we also note that the problem size increased 16 times. The
mean time increases about ten times as we increase the network size by
a factor of two. For $\gamma = 10$, the mean time increases about
five times as the number of units increases twice (from 160 to 320,
and from 320 to 640 units). Interestingly, the time for $\gamma_p =
10 \%$ is roughly the same for 160, 320, and 640 units, which suggests
the time complexity depends on the number of units requested relative
to the number of available units, and indirectly on the spectrum
fragmentation. The mean time decreases as the network utilization
increases, since the search space gets smaller. As for the sample
maximum results, they were usually a hundred times larger than the
mean results.

% The memory.

The memory results report the number of 64-bit memory words used by
the permanent solutions, the tentative solutions, and the priority
queue. The network size, $|\Omega|$, and $\gamma$ affected the memory
results similar to how they affected the time results. For the
networks with 25 nodes, the mean number of words was about $10^5$,
while, for the networks with 100~nodes, about $10^9$.

% The difference between $\gamma = 10$ and $\gamma_p = 10 \%$

The memory used for $\gamma = 10$ is far larger than for $\gamma_p =
10 \%$ because the spectrum (the available units) is more fragmented
(since it is allocated in smaller fragments), and the algorithm finds
more solutions as the search space is larger. Finding more solutions
requires more time: simulations for $\gamma = 10$ take more time that
the simulations for $\gamma_p = 10 \%$.

% The memory used: the stack plots.

To examine how the incomparable permanent and tentative labels, and
the elements of the priority queue contribute to the memory usage,
Figure~\ref{f:stack} shows as stack plots the maximal memory used by the
proposed algorithm for the networks of 25, 50, and 100~nodes with 320
units, and $\gamma = 10$.

% The memory used: the explanation.

The permanent labels take about 80\% of memory, the tentative labels
about 20\%, and the elements of the priority queue take only a small
fraction. We assumed that a label takes 15 64-bit words
(implementation details: 4 words for a shared pointer, 1 word for a
vertex pair, 4 words for a path trait, 2 words for an edge, 4 words
for a shared pointer to the parent node in the search tree). An
element of a priority queue is two words long (1 word for cost, 1 word
for a weak pointer to its tentative label).

% The memory used: the conclusion.

Most of the memory required by the algorithm is consumed by the
permanent labels because of the large search space. A permanent
label stores an optimal solution, and the results suggest that there
are many of them for large networks. Furthermore, that large number
of permanent labels helps to keep the number of tentative labels
relatively much smaller through the edge relaxation.

% BBP: the figure

To further validate the proposed algorithm, Figure~\ref{f:bbp} shows,
for all populations of interest, the mean bandwidth blocking
probabilities of the proposed algorithm as thin curves and of the
edge-exclusion algorithm as thick curves. The figure has two rows and
three columns of subfigures. The first row shows the results for
$\gamma = 10$, and the second for $\gamma_p = 10 \%$. The first
column shows the results for the networks with 25 nodes, the second
for 50 nodes, and the third for 100 nodes. We do not plot the error
bars representing the standard error, since they were too small to
plot.

\vspace{-6pt}

% start a new page without indent 4.6cm
%\clearpage
\end{paracol}
\nointerlineskip
\begin{figure}[H]\widefigure  \vspace{-6pt} 
  \subfloat[25 node networks.]{%
    \label{f:stack_dijkstra}%
    \input{Definitions/stack_25}}\hfill%
  \subfloat[50 node networks.]{%
    \label{f:stack_parallel}%
    \input{Definitions/stack_50}}\hfill%
  \subfloat[100 node networks.]{%
    \label{f:stack_brtforce}%
    \input{Definitions/stack_100}}\\[5pt]%
  \centering\ref{stack}%
  \caption{Simulation results: the maximum number of required words
 for networks with 320 units, and $\gamma = 10$.}
  \label{f:stack}
\end{figure}
\begin{paracol}{2}
%\linenumbers
\switchcolumn

\end{paracol}
\nointerlineskip

\begin{figure}[H]\widefigure
  \subfloat[memory used for $\gamma = 10$, and 25 nodes]{%
    \label{f:bbp_10_25}%
    \input{Definitions/bbp_10_25}}\hfill%
  \subfloat[memory used for $\gamma_p = 10\%$, and 25 nodes]{%
    \label{f:bbp_10p_25}%
    \input{Definitions/bbp_10p_25}}\\[5pt]%
  \subfloat[memory used for $\gamma = 10$, and 50 nodes]{%
    \label{f:bbp_10_50}%
    \input{Definitions/bbp_10_50}}\hfill%
  \subfloat[memory used for $\gamma_p = 10\%$, and 50 nodes]{%
    \label{f:bbp_10p_50}%
    \input{Definitions/bbp_10p_50}}\\[5pt]%
  \subfloat[memory used for $\gamma = 10$, and 100 nodes]{%
    \label{f:bbp_10_100}%
    \input{Definitions/bbp_10_100}}\hfill%
  \subfloat[memory used for $\gamma_p = 10\%$, and 100 nodes]{%
    \label{f:bbp_10p_100}%
    \input{Definitions/bbp_10p_100}}\\[5pt]%
  \centering\ref{bbp}%
  \caption{Simulation results: the sample means of the bandwidth
 blocking probability.}
  \label{f:bbp}
\end{figure}
% start a new page without indent 4.6cm

\begin{paracol}{2}
%\linenumbers
\switchcolumn

Since the proposed algorithm can be exact, and the edge-exclusion
algorithm is heuristic, the proposed algorithm should perform better,
and indeed this is so. Interestingly, the edge-exclusion algorithm
(which uses the generic Dijkstra algorithm) performs very well, at only
about 5\% worse.

% Comparison: edge-exclusion and brute-force not there.

We did not add the edge-exclusion algorithm to the time and memory
performance comparison, since it is a heuristic algorithm with the
worst-case computational complexity of the dynamic routing problem
without DPP, i.e., $O(|\Omega|^2|V|log|V|)$
\cite{10.1364/JOCN.11.000568}. In addition, we were unable to add the
brute-force algorithm to the comparison because of its exponential
complexity.

% Let's put the results in perspective.

Admittedly, the reported time and memory consumption of the proposed
algorithm seems large: for a network of 100 nodes and 640 units, the
algorithm can run even for thousands of seconds, and use even 10 GBs
of operating memory. However, to put these results in perspective, we
report that, for a \emph{far smaller} network of 15 nodes and 640
units, the brute-force algorithm ran for thousands of seconds, and it
requested more than 120 GBs of operating memory in the corroborative
runs.

%%%%%%%%%%%%%%%%%%%%%%%%%%%%%%%%%%%%%%%%%%%%%%%%%%%%%%%%%%%%%%%%%%%%%%%%%%%

\section{Conclusions}
\label{conclusion}

% A step in the right direction, the basis.

The proposed algorithm is capable of solving various dynamic routing
problems with dedicated path protection in optical networks, but not
all of them, e.g., the algorithm cannot be applied when signal
regeneration or spectrum conversion are used. However, the proposed
algorithm can solve those routing problems that meet the minimal
requirements of the stated research problem. If not, then perhaps the
proposed algorithm and its novel principles could be used as a basis
for devising more capable algorithms.

% Other uses.

The proposed algorithm can also be used to find a pair of paths to
different (primary, and secondary) data centers. The algorithm could
even be useful in routing with inverse multiplexing, and multipath
routing.

% The performance comparison.

We are unable to compare the performance results of the proposed
algorithm to some efficient and exact algorithm because, to the best
of our knowledge, \emph{no such competing algorithm exists}. For that
large problem size, %Please ensure the meaning has been retained.  Author: It's OK.
we could not have used the existing exact methods
(e.g., the brute-force algorithm, integer linear programming) since
they are inefficient, nor could we have used the existing efficient
methods (e.g., the edge-exclusion algorithm, the tabu search), since
they are suboptimal.

% Different layers could be protected.

Dedicated path protection can be implemented at the multiplex section
(fiber protection), the optical layer (optical signal protection), or
the digital layer (the digital signal protection). We presented our
algorithm in the setting of the optical signal protection, where we
take into account the spectrum continuity and contiguity constraints,
but the same principles could be used for other layers (e.g., the
Internet protocol layer with multiprotocol label switching) and
networks, too.

% Adaptations.

The algorithm can be adapted for further constraints, e.g.,
node-disjoint paths or the same spectrum fragments on both paths.
Node-disjoint paths can be found if we do not relax the edges of the
search graph that leave vertex $(v_{x, 1}, v_{x, 2})$ when $v_{x, 1}
== v_{x, 2}$. The same spectrum fragment on both paths can be
enforced by making sure during edge relaxation that the intersection
of two fragments meet the requirements of a demand.

% Future work.

Future work could concentrate on applying the algorithm to related
problems (e.g., establishing protected content-oriented connections to
data centers), and further improving its performance with parallel
computing.

% Suurballe algorithm and the incomparability principle.

Furthermore, perhaps the principle of the incomparability of solutions
could be applied to the path augmentation technique, thus making,
e.g., the Suurballe algorithm, even faster than the proposed algorithm.

% The implementation.

The provided implementation does not require proprietary software, is
implemented in modern C++ using the Boost Graph Library, and with
modern functionality, such as concepts, smart pointers, in-place object
creation, and move semantics. The implementation can be used to
replicate the presented results, as well as stress-test the proposed
algorithm.

\clearpage
\authorcontributions{Conceptualization, I.S. and I.O.; methodology, I.S., I.O, and B.WS.; software, I.S.; validation, I.S., I.O., and B.WS., writing---original draft preparation, I.S.; writing---review and editing, I.S., I.O, and B.WS.. All authors have read and agreed to the published version of the manuscript.}

\funding{This research was funded by the Polish Ministry of Science
  and Higher Education grant number 020/RID/2018/19.}

\institutionalreview{Not applicable.}

\informedconsent{Not applicable.}

%MDPI: Please ensure that all individuals included in this section have consented to the acknowledgement.  Author: ensured.
\acknowledgments{We ran the simulations using PL-Grid, the Polish
  supercomputing infrastructure. %MDPI: In this section you can acknowledge any support given which is not covered by the author contribution or funding sections. This may include administrative and technical support, or donations in kind (e.g., materials used for experiments).
}

\conflictsofinterest{The authors declare no conflict of interest.}

\end{paracol}

\reftitle{References}

\end{document}

%% file: Definitions/time_10_25.tex
\begin{tikzpicture}
\begin{semilogyaxis}[xlabel = {network utilization}, ylabel = {time [s]}, height = 6.5 cm, width = 8.5 cm, grid = major, ticks = major, legend columns = 3, legend to name = regular, legend style = {/tikz/every even column/.append style = {column sep = 0.25 cm}}, xmin = 0.1, xmax = 0.5, ymin = 1e-4, ymax = 1e5]
\addplot[dotted]
coordinates {
(0.143942, 0.0018253) +- (0.00910836, 0.000162545)
(0.206649, 0.00235859) +- (0.00836592, 0.00019914)
(0.246022, 0.0023091) +- (0.00812366, 0.000195287)
(0.271731, 0.00208456) +- (0.00811458, 0.000148851)
(0.336609, 0.00156858) +- (0.00746947, 9.57089e-05)
(0.355467, 0.0013725) +- (0.00720303, 8.63295e-05)
(0.379126, 0.00107231) +- (0.00751296, 5.68827e-05)
(0.400608, 0.000957327) +- (0.00661102, 5.24588e-05)
(0.415544, 0.000833453) +- (0.00650509, 3.95442e-05)
};
\addlegendentry{mean, 160 units}
\addplot[dashed]
coordinates {
(0.145098, 0.00976868) +- (0.00629342, 0.00126819)
(0.238705, 0.0112174) +- (0.00823697, 0.00114459)
(0.284467, 0.00973288) +- (0.00814058, 0.000978314)
(0.310308, 0.00882759) +- (0.00780967, 0.000818867)
(0.36885, 0.00566993) +- (0.00747156, 0.000384903)
(0.389151, 0.00507383) +- (0.00736526, 0.00036753)
(0.413018, 0.00412616) +- (0.00704032, 0.000250908)
(0.43612, 0.00336573) +- (0.00729952, 0.000192361)
(0.452103, 0.00277924) +- (0.00710904, 0.000140217)
};
\addlegendentry{mean, 320 units}
\addplot[solid]
coordinates {
(0.154797, 0.074471) +- (0.00706633, 0.0111855)
(0.261136, 0.0880417) +- (0.00852862, 0.013083)
(0.308136, 0.0642631) +- (0.00773671, 0.00875143)
(0.338818, 0.0479205) +- (0.00808019, 0.0050533)
(0.396532, 0.0318169) +- (0.00752232, 0.00286835)
(0.419071, 0.0256454) +- (0.00731279, 0.00220301)
(0.442737, 0.0202313) +- (0.00752019, 0.00152789)
(0.467174, 0.0159064) +- (0.00753052, 0.00113451)
(0.481076, 0.013749) +- (0.00752028, 0.000858182)
};
\addlegendentry{mean, 640 units}
\addplot[thick, dotted]
coordinates {
(0.143942, 0.0516813) +- (0.00910836, 0)
(0.206649, 0.0786038) +- (0.00836592, 0)
(0.246022, 0.123926) +- (0.00812366, 0)
(0.271731, 0.115088) +- (0.00811458, 0)
(0.336609, 0.0979301) +- (0.00746947, 0)
(0.355467, 0.268272) +- (0.00720303, 0)
(0.379126, 0.193834) +- (0.00751296, 0)
(0.400608, 0.152121) +- (0.00661102, 0)
(0.415544, 0.176197) +- (0.00650509, 0)
};
\addlegendentry{maximum, 160 units}
\addplot[thick, dashed]
coordinates {
(0.145098, 0.654656) +- (0.00629342, 0)
(0.238705, 1.15424) +- (0.00823697, 0)
(0.284467, 1.30491) +- (0.00814058, 0)
(0.310308, 0.775457) +- (0.00780967, 0)
(0.36885, 1.13936) +- (0.00747156, 0)
(0.389151, 1.10818) +- (0.00736526, 0)
(0.413018, 1.23889) +- (0.00704032, 0)
(0.43612, 0.970656) +- (0.00729952, 0)
(0.452103, 0.546787) +- (0.00710904, 0)
};
\addlegendentry{maximum, 320 units}
\addplot[thick, solid]
coordinates {
(0.154797, 5.597) +- (0.00706633, 0)
(0.261136, 13.6621) +- (0.00852862, 0)
(0.308136, 21.5537) +- (0.00773671, 0)
(0.338818, 7.90935) +- (0.00808019, 0)
(0.396532, 15.3913) +- (0.00752232, 0)
(0.419071, 17.836) +- (0.00731279, 0)
(0.442737, 11.5338) +- (0.00752019, 0)
(0.467174, 16.5737) +- (0.00753052, 0)
(0.481076, 21.735) +- (0.00752028, 0)
};
\addlegendentry{maximum, 640 units}
\end{semilogyaxis}
\end{tikzpicture}

%% file: Definitions/time_10p_25.tex
\begin{tikzpicture}
\begin{semilogyaxis}[xlabel = {network utilization}, ylabel = {time [s]}, height = 6.5 cm, width = 8.5 cm, grid = major, ticks = major, legend columns = 3, legend to name = regular, legend style = {/tikz/every even column/.append style = {column sep = 0.25 cm}}, xmin = 0.075, xmax = 0.4, ymin = 1e-4, ymax = 1e3]
\addplot[dotted]
coordinates {
(0.113935, 0.000941489) +- (0.00713754, 6.89987e-05)
(0.181142, 0.00101662) +- (0.00732259, 6.9656e-05)
(0.225823, 0.000997728) +- (0.00806008, 8.3509e-05)
(0.256669, 0.000858708) +- (0.00708683, 6.0845e-05)
(0.305974, 0.000661648) +- (0.00790537, 4.17127e-05)
(0.326675, 0.000584836) +- (0.00689427, 3.14819e-05)
(0.353531, 0.000473101) +- (0.00668977, 2.22339e-05)
(0.375944, 0.000397221) +- (0.00702454, 1.79995e-05)
(0.390798, 0.000341578) +- (0.00688805, 1.43158e-05)
};
\addlegendentry{mean, 160 units}
\addplot[dashed]
coordinates {
(0.103148, 0.000873279) +- (0.00672575, 6.19771e-05)
(0.180577, 0.000996381) +- (0.00648534, 7.57143e-05)
(0.224838, 0.000886967) +- (0.00759845, 6.03022e-05)
(0.254381, 0.000846762) +- (0.00727339, 6.47396e-05)
(0.311084, 0.000598968) +- (0.0076647, 3.72966e-05)
(0.328403, 0.000517609) +- (0.00781389, 2.42393e-05)
(0.354752, 0.000432701) +- (0.00712736, 2.03612e-05)
(0.375796, 0.000358623) +- (0.00732195, 1.44424e-05)
(0.391025, 0.000316662) +- (0.00665829, 1.41769e-05)
};
\addlegendentry{mean, 320 units}
\addplot[solid]
coordinates {
(0.107152, 0.00088487) +- (0.00597721, 5.85318e-05)
(0.177023, 0.000983345) +- (0.00679407, 7.31064e-05)
(0.22269, 0.000898255) +- (0.00759882, 6.5571e-05)
(0.257486, 0.000798733) +- (0.00785466, 5.69244e-05)
(0.311219, 0.000580519) +- (0.0074807, 3.49436e-05)
(0.331221, 0.000510894) +- (0.00729832, 2.73674e-05)
(0.354567, 0.000420931) +- (0.00707648, 2.08485e-05)
(0.376776, 0.000337349) +- (0.00730113, 1.41278e-05)
(0.391779, 0.000293258) +- (0.00706304, 1.13078e-05)
};
\addlegendentry{mean, 640 units}
\addplot[thick, dotted]
coordinates {
(0.113935, 0.0223279) +- (0.00713754, 0)
(0.181142, 0.0234287) +- (0.00732259, 0)
(0.225823, 0.0566186) +- (0.00806008, 0)
(0.256669, 0.0261975) +- (0.00708683, 0)
(0.305974, 0.0361918) +- (0.00790537, 0)
(0.326675, 0.0556885) +- (0.00689427, 0)
(0.353531, 0.0179799) +- (0.00668977, 0)
(0.375944, 0.0219305) +- (0.00702454, 0)
(0.390798, 0.0234216) +- (0.00688805, 0)
};
\addlegendentry{maximum, 160 units}
\addplot[thick, dashed]
coordinates {
(0.103148, 0.0121288) +- (0.00672575, 0)
(0.180577, 0.019786) +- (0.00648534, 0)
(0.224838, 0.0275509) +- (0.00759845, 0)
(0.254381, 0.0387744) +- (0.00727339, 0)
(0.311084, 0.0362556) +- (0.0076647, 0)
(0.328403, 0.0151475) +- (0.00781389, 0)
(0.354752, 0.013526) +- (0.00712736, 0)
(0.375796, 0.0128401) +- (0.00732195, 0)
(0.391025, 0.0165861) +- (0.00665829, 0)
};
\addlegendentry{maximum, 320 units}
\addplot[thick, solid]
coordinates {
(0.107152, 0.0147075) +- (0.00597721, 0)
(0.177023, 0.0205201) +- (0.00679407, 0)
(0.22269, 0.0180568) +- (0.00759882, 0)
(0.257486, 0.0350139) +- (0.00785466, 0)
(0.311219, 0.0241171) +- (0.0074807, 0)
(0.331221, 0.0136029) +- (0.00729832, 0)
(0.354567, 0.0114477) +- (0.00707648, 0)
(0.376776, 0.0218404) +- (0.00730113, 0)
(0.391779, 0.0074429) +- (0.00706304, 0)
};
\addlegendentry{maximum, 640 units}
\end{semilogyaxis}
\end{tikzpicture}

%% file: Definitions/time_10_50.tex
\begin{tikzpicture}
\begin{semilogyaxis}[xlabel = {network utilization}, ylabel = {time [s]}, height = 6.5 cm, width = 8.5 cm, grid = major, ticks = major, legend columns = 3, legend to name = regular, legend style = {/tikz/every even column/.append style = {column sep = 0.25 cm}}, xmin = 0.1, xmax = 0.5, ymin = 1e-4, ymax = 1e5]
\addplot[dotted]
coordinates {
(0.137347, 0.0269846) +- (0.00417829, 0.00242421)
(0.209375, 0.0306643) +- (0.00506164, 0.00300472)
(0.246112, 0.0260806) +- (0.00501929, 0.00246235)
(0.268306, 0.0243366) +- (0.00549038, 0.00241634)
(0.314408, 0.0154952) +- (0.0049428, 0.00121364)
(0.334647, 0.0121032) +- (0.00497782, 0.000930967)
(0.351075, 0.0105708) +- (0.00485923, 0.000959472)
(0.370752, 0.00762092) +- (0.00458584, 0.000472532)
(0.386705, 0.00625212) +- (0.00466367, 0.000409199)
};
\addlegendentry{mean, 160 units}
\addplot[dashed]
coordinates {
(0.143737, 0.235061) +- (0.00423325, 0.0287048)
(0.239671, 0.229817) +- (0.00541046, 0.0345519)
(0.276411, 0.169363) +- (0.00502445, 0.0235854)
(0.301211, 0.12638) +- (0.00473804, 0.0147022)
(0.345596, 0.0744568) +- (0.00477823, 0.00798615)
(0.362728, 0.0547439) +- (0.00498939, 0.00457205)
(0.383287, 0.0414837) +- (0.00486709, 0.00353582)
(0.403796, 0.0276863) +- (0.0046782, 0.00195485)
(0.418844, 0.0234244) +- (0.0045363, 0.00166915)
};
\addlegendentry{mean, 320 units}
\addplot[solid]
coordinates {
(0.156029, 2.97968) +- (0.00508507, 0.524362)
(0.245737, 2.34628) +- (0.00482973, 0.444846)
(0.296806, 1.22244) +- (0.00492433, 0.180331)
(0.321653, 0.962175) +- (0.00455477, 0.154905)
(0.36933, 0.424312) +- (0.00514214, 0.0477417)
(0.387559, 0.310498) +- (0.00519268, 0.0320401)
(0.409804, 0.226163) +- (0.0048596, 0.0227525)
(0.428943, 0.167353) +- (0.00481172, 0.0157515)
(0.443752, 0.137488) +- (0.00508116, 0.0103961)
};
\addlegendentry{mean, 640 units}
\addplot[thick, dotted]
coordinates {
(0.137347, 1.79697) +- (0.00417829, 0)
(0.209375, 2.5866) +- (0.00506164, 0)
(0.246112, 2.24188) +- (0.00501929, 0)
(0.268306, 3.63083) +- (0.00549038, 0)
(0.314408, 3.27822) +- (0.0049428, 0)
(0.334647, 5.63195) +- (0.00497782, 0)
(0.351075, 21.0062) +- (0.00485923, 0)
(0.370752, 4.04652) +- (0.00458584, 0)
(0.386705, 8.41444) +- (0.00466367, 0)
};
\addlegendentry{maximum, 160 units}
\addplot[thick, dashed]
coordinates {
(0.143737, 24.1062) +- (0.00423325, 0)
(0.239671, 87.3324) +- (0.00541046, 0)
(0.276411, 80.8715) +- (0.00502445, 0)
(0.301211, 50.5474) +- (0.00473804, 0)
(0.345596, 199.415) +- (0.00477823, 0)
(0.362728, 53.5652) +- (0.00498939, 0)
(0.383287, 84.1718) +- (0.00486709, 0)
(0.403796, 57.7272) +- (0.0046782, 0)
(0.418844, 59.0666) +- (0.0045363, 0)
};
\addlegendentry{maximum, 320 units}
\addplot[thick, solid]
coordinates {
(0.156029, 628.214) +- (0.00508507, 0)
(0.245737, 788.301) +- (0.00482973, 0)
(0.296806, 982.475) +- (0.00492433, 0)
(0.321653, 914.459) +- (0.00455477, 0)
(0.36933, 555.311) +- (0.00514214, 0)
(0.387559, 778.806) +- (0.00519268, 0)
(0.409804, 1021.68) +- (0.0048596, 0)
(0.428943, 1173.96) +- (0.00481172, 0)
(0.443752, 736.048) +- (0.00508116, 0)
};
\addlegendentry{maximum, 640 units}
\end{semilogyaxis}
\end{tikzpicture}

%% file: Definitions/time_10p_50.tex
\begin{tikzpicture}
\begin{semilogyaxis}[xlabel = {network utilization}, ylabel = {time [s]}, height = 6.5 cm, width = 8.5 cm, grid = major, ticks = major, legend columns = 3, legend to name = regular, legend style = {/tikz/every even column/.append style = {column sep = 0.25 cm}}, xmin = 0.075, xmax = 0.4, ymin = 1e-4, ymax = 1e3]
\addplot[dotted]
coordinates {
(0.117661, 0.0101142) +- (0.00569911, 0.000871581)
(0.188487, 0.0104505) +- (0.00599344, 0.000870952)
(0.220736, 0.00976013) +- (0.0059305, 0.000852781)
(0.247075, 0.00841695) +- (0.00565563, 0.000698991)
(0.291796, 0.0056311) +- (0.00524972, 0.000391837)
(0.31307, 0.0044655) +- (0.00579297, 0.000275168)
(0.332693, 0.00330867) +- (0.00505063, 0.000178842)
(0.351376, 0.00261922) +- (0.00446423, 0.000144286)
(0.362726, 0.00217072) +- (0.00483861, 9.945e-05)
};
\addlegendentry{mean, 160 units}
\addplot[dashed]
coordinates {
(0.102604, 0.00853994) +- (0.00513301, 0.000770839)
(0.181763, 0.00951806) +- (0.00564969, 0.000770322)
(0.226664, 0.00848468) +- (0.00565136, 0.000602797)
(0.244222, 0.00765555) +- (0.00534454, 0.000634919)
(0.297387, 0.00467039) +- (0.004939, 0.000290458)
(0.31379, 0.00371983) +- (0.00565221, 0.000212133)
(0.334351, 0.00287494) +- (0.00489876, 0.000152324)
(0.350942, 0.00217531) +- (0.00452578, 9.8099e-05)
(0.364262, 0.00179228) +- (0.00494357, 7.52379e-05)
};
\addlegendentry{mean, 320 units}
\addplot[solid]
coordinates {
(0.0966696, 0.00739783) +- (0.00420626, 0.000561434)
(0.188585, 0.00935505) +- (0.00615896, 0.000747555)
(0.228492, 0.00831721) +- (0.005742, 0.000649744)
(0.249141, 0.00706737) +- (0.00559674, 0.000582924)
(0.293901, 0.00455305) +- (0.00538317, 0.000289156)
(0.3112, 0.0036705) +- (0.00549271, 0.000213321)
(0.331432, 0.00267237) +- (0.00564668, 0.000142934)
(0.349545, 0.00205272) +- (0.00518425, 0.000100409)
(0.364471, 0.00162459) +- (0.00502601, 6.81315e-05)
};
\addlegendentry{mean, 640 units}
\addplot[thick, dotted]
coordinates {
(0.117661, 0.283775) +- (0.00569911, 0)
(0.188487, 0.478618) +- (0.00599344, 0)
(0.220736, 0.971531) +- (0.0059305, 0)
(0.247075, 0.947866) +- (0.00565563, 0)
(0.291796, 0.468475) +- (0.00524972, 0)
(0.31307, 0.780555) +- (0.00579297, 0)
(0.332693, 0.762389) +- (0.00505063, 0)
(0.351376, 0.567269) +- (0.00446423, 0)
(0.362726, 0.366911) +- (0.00483861, 0)
};
\addlegendentry{maximum, 160 units}
\addplot[thick, dashed]
coordinates {
(0.102604, 0.196156) +- (0.00513301, 0)
(0.181763, 0.303941) +- (0.00564969, 0)
(0.226664, 0.376854) +- (0.00565136, 0)
(0.244222, 0.520715) +- (0.00534454, 0)
(0.297387, 0.567415) +- (0.004939, 0)
(0.31379, 0.472803) +- (0.00565221, 0)
(0.334351, 0.63933) +- (0.00489876, 0)
(0.350942, 0.194559) +- (0.00452578, 0)
(0.364262, 0.177096) +- (0.00494357, 0)
};
\addlegendentry{maximum, 320 units}
\addplot[thick, solid]
coordinates {
(0.0966696, 0.199764) +- (0.00420626, 0)
(0.188585, 0.308197) +- (0.00615896, 0)
(0.228492, 0.670132) +- (0.005742, 0)
(0.249141, 0.439841) +- (0.00559674, 0)
(0.293901, 0.25887) +- (0.00538317, 0)
(0.3112, 0.402433) +- (0.00549271, 0)
(0.331432, 0.213947) +- (0.00564668, 0)
(0.349545, 0.0984935) +- (0.00518425, 0)
(0.364471, 0.106758) +- (0.00502601, 0)
};
\addlegendentry{maximum, 640 units}
\end{semilogyaxis}
\end{tikzpicture}

%% file: Definitions/time_10_100.tex
\begin{tikzpicture}
\begin{semilogyaxis}[xlabel = {network utilization}, ylabel = {time [s]}, height = 6.5 cm, width = 8.5 cm, grid = major, ticks = major, legend columns = 3, legend to name = regular, legend style = {/tikz/every even column/.append style = {column sep = 0.25 cm}}, xmin = 0.1, xmax = 0.5, ymin = 1e-4, ymax = 1e5]
\addplot[dotted]
coordinates {
(0.138656, 0.60263) +- (0.00349105, 0.0731372)
(0.217396, 0.565873) +- (0.00338113, 0.0672728)
(0.250139, 0.461208) +- (0.00342693, 0.0488296)
(0.265751, 0.404694) +- (0.00358388, 0.0405974)
(0.304259, 0.216922) +- (0.00306796, 0.0218984)
(0.314785, 0.180405) +- (0.00331254, 0.016222)
(0.333959, 0.121502) +- (0.00298775, 0.00936493)
(0.350223, 0.0869936) +- (0.00274093, 0.00640906)
(0.361814, 0.0624728) +- (0.00292915, 0.00409972)
};
\addlegendentry{mean, 160 units}
\addplot[dashed]
coordinates {
(0.158062, 7.13288) +- (0.00276868, 0.935109)
(0.2369, 5.07941) +- (0.00294746, 0.726882)
(0.271127, 2.93572) +- (0.00297365, 0.330087)
(0.291183, 2.13708) +- (0.00290559, 0.253732)
(0.327215, 1.0995) +- (0.00306662, 0.105938)
(0.341482, 0.858156) +- (0.0030209, 0.0939888)
(0.358734, 0.52705) +- (0.00307359, 0.0475713)
(0.375301, 0.397071) +- (0.00269418, 0.041134)
(0.386229, 0.295967) +- (0.00275794, 0.0227792)
};
\addlegendentry{mean, 320 units}
\addplot[solid]
coordinates {
(0.148295, 105.467) +- (0.00295821, 13.5112)
(0.238104, 50.5732) +- (0.00267535, 7.84394)
(0.284313, 26.73) +- (0.00252097, 3.64748)
(0.307893, 16.5458) +- (0.00301953, 2.04524)
(0.347749, 6.54288) +- (0.00321284, 0.629938)
(0.363006, 5.0434) +- (0.00312716, 0.508517)
(0.380794, 3.35837) +- (0.00298298, 0.325007)
(0.397339, 2.55412) +- (0.0029367, 0.250528)
(0.409587, 1.93105) +- (0.00287135, 0.16111)
};
\addlegendentry{mean, 640 units}
\addplot[thick, dotted]
coordinates {
(0.138656, 133.77) +- (0.00349105, 0)
(0.217396, 267.403) +- (0.00338113, 0)
(0.250139, 133.457) +- (0.00342693, 0)
(0.265751, 123.039) +- (0.00358388, 0)
(0.304259, 499.045) +- (0.00306796, 0)
(0.314785, 370.376) +- (0.00331254, 0)
(0.333959, 192.96) +- (0.00298775, 0)
(0.350223, 144.545) +- (0.00274093, 0)
(0.361814, 108.602) +- (0.00292915, 0)
};
\addlegendentry{maximum, 160 units}
\addplot[thick, dashed]
coordinates {
(0.158062, 1053.71) +- (0.00276868, 0)
(0.2369, 6408.01) +- (0.00294746, 0)
(0.271127, 1489.86) +- (0.00297365, 0)
(0.291183, 1820.75) +- (0.00290559, 0)
(0.327215, 1271.68) +- (0.00306662, 0)
(0.341482, 2231.43) +- (0.0030209, 0)
(0.358734, 2346.34) +- (0.00307359, 0)
(0.375301, 3788.51) +- (0.00269418, 0)
(0.386229, 1365.22) +- (0.00275794, 0)
};
\addlegendentry{maximum, 320 units}
\addplot[thick, solid]
coordinates {
(0.148295, 17291.5) +- (0.00295821, 0)
(0.238104, 33419.8) +- (0.00267535, 0)
(0.284313, 42779.2) +- (0.00252097, 0)
(0.307893, 29391.9) +- (0.00301953, 0)
(0.347749, 11538.7) +- (0.00321284, 0)
(0.363006, 18478.4) +- (0.00312716, 0)
(0.380794, 39855.6) +- (0.00298298, 0)
(0.397339, 93506.7) +- (0.0029367, 0)
(0.409587, 47498.4) +- (0.00287135, 0)
};
\addlegendentry{maximum, 640 units}
\end{semilogyaxis}
\end{tikzpicture}

%% file: Definitions/time_10p_100.tex
\begin{tikzpicture}
\begin{semilogyaxis}[xlabel = {network utilization}, ylabel = {time [s]}, height = 6.5 cm, width = 8.5 cm, grid = major, ticks = major, legend columns = 3, legend to name = regular, legend style = {/tikz/every even column/.append style = {column sep = 0.25 cm}}, xmin = 0.075, xmax = 0.4, ymin = 1e-4, ymax = 1e3]
\addplot[dotted]
coordinates {
(0.119977, 0.142553) +- (0.00348355, 0.0120872)
(0.190231, 0.157888) +- (0.0039338, 0.0161484)
(0.226321, 0.134019) +- (0.00411252, 0.0123548)
(0.243865, 0.114465) +- (0.00379688, 0.00933081)
(0.288175, 0.0609595) +- (0.00338089, 0.00453749)
(0.299012, 0.0482725) +- (0.00349773, 0.00322694)
(0.316203, 0.0329339) +- (0.00346146, 0.00214667)
(0.329426, 0.0238073) +- (0.00298566, 0.00131265)
(0.341588, 0.0176823) +- (0.00293003, 0.000896137)
};
\addlegendentry{mean, 160 units}
\addplot[dashed]
coordinates {
(0.11443, 0.130807) +- (0.00335236, 0.011976)
(0.188332, 0.141484) +- (0.0038314, 0.0107629)
(0.226132, 0.11657) +- (0.00412051, 0.0110518)
(0.248117, 0.0992952) +- (0.00396907, 0.01002)
(0.288263, 0.0511533) +- (0.00365752, 0.00381608)
(0.301603, 0.0367482) +- (0.0034388, 0.00240992)
(0.315909, 0.0261824) +- (0.00315253, 0.00138822)
(0.330948, 0.0182582) +- (0.00306049, 0.000930548)
(0.342636, 0.0139189) +- (0.00289001, 0.000678135)
};
\addlegendentry{mean, 320 units}
\addplot[solid]
coordinates {
(0.103775, 0.13016) +- (0.00298092, 0.0103799)
(0.197487, 0.148049) +- (0.00356882, 0.0139691)
(0.231169, 0.105878) +- (0.00336791, 0.00975879)
(0.247417, 0.090528) +- (0.0037526, 0.00693265)
(0.288076, 0.045841) +- (0.00345113, 0.0032391)
(0.303308, 0.0330248) +- (0.00356083, 0.00197552)
(0.31731, 0.02213) +- (0.00361482, 0.00116177)
(0.331111, 0.0154521) +- (0.00330166, 0.000730609)
(0.341973, 0.0122505) +- (0.00287774, 0.000606415)
};
\addlegendentry{mean, 640 units}
\addplot[thick, dotted]
coordinates {
(0.119977, 11.4266) +- (0.00348355, 0)
(0.190231, 20.2038) +- (0.0039338, 0)
(0.226321, 21.0907) +- (0.00411252, 0)
(0.243865, 13.8413) +- (0.00379688, 0)
(0.288175, 32.1903) +- (0.00338089, 0)
(0.299012, 18.3759) +- (0.00349773, 0)
(0.316203, 18.766) +- (0.00346146, 0)
(0.329426, 9.72698) +- (0.00298566, 0)
(0.341588, 7.66788) +- (0.00293003, 0)
};
\addlegendentry{maximum, 160 units}
\addplot[thick, dashed]
coordinates {
(0.11443, 8.44489) +- (0.00335236, 0)
(0.188332, 9.01283) +- (0.0038314, 0)
(0.226132, 10.4348) +- (0.00412051, 0)
(0.248117, 18.7578) +- (0.00396907, 0)
(0.288263, 11.7729) +- (0.00365752, 0)
(0.301603, 12.1725) +- (0.0034388, 0)
(0.315909, 5.91404) +- (0.00315253, 0)
(0.330948, 5.049) +- (0.00306049, 0)
(0.342636, 4.89365) +- (0.00289001, 0)
};
\addlegendentry{maximum, 320 units}
\addplot[thick, solid]
coordinates {
(0.103775, 5.36326) +- (0.00298092, 0)
(0.197487, 21.9753) +- (0.00356882, 0)
(0.231169, 6.73652) +- (0.00336791, 0)
(0.247417, 8.05432) +- (0.0037526, 0)
(0.288076, 11.0337) +- (0.00345113, 0)
(0.303308, 8.85718) +- (0.00356083, 0)
(0.31731, 3.43515) +- (0.00361482, 0)
(0.331111, 1.75375) +- (0.00330166, 0)
(0.341973, 2.30116) +- (0.00287774, 0)
};
\addlegendentry{maximum, 640 units}
\end{semilogyaxis}
\end{tikzpicture}

%% file: Definitions/memory_10_25.tex
\begin{tikzpicture}
\begin{semilogyaxis}[xlabel = {network utilization}, ylabel = {memory words}, height = 6.5 cm, width = 8.5 cm, grid = major, ticks = major, legend columns = 3, legend to name = regular, legend style = {/tikz/every even column/.append style = {column sep = 0.25 cm}}, xmin = 0.1, xmax = 0.5, ymin = 1e3, ymax = 1e10]
\addplot[dotted]
coordinates {
(0.143942, 15146.2) +- (0.00910836, 1022.79)
(0.206649, 17857.0) +- (0.00836592, 1086.85)
(0.246022, 16987.0) +- (0.00812366, 992.689)
(0.271731, 15355.2) +- (0.00811458, 761.694)
(0.336609, 11468.7) +- (0.00746947, 507.727)
(0.355467, 9912.69) +- (0.00720303, 419.811)
(0.379126, 7809.69) +- (0.00751296, 285.78)
(0.400608, 6827.23) +- (0.00661102, 261.925)
(0.415544, 5961.09) +- (0.00650509, 205.394)
};
\addlegendentry{mean, 160 units}
\addplot[dashed]
coordinates {
(0.145098, 51716.0) +- (0.00629342, 3768.45)
(0.238705, 56323.9) +- (0.00823697, 3577.5)
(0.284467, 49389.4) +- (0.00814058, 2938.92)
(0.310308, 45463.7) +- (0.00780967, 2662.82)
(0.36885, 30548.0) +- (0.00747156, 1252.5)
(0.389151, 26806.5) +- (0.00736526, 1120.89)
(0.413018, 22216.5) +- (0.00704032, 812.332)
(0.43612, 18155.0) +- (0.00729952, 634.827)
(0.452103, 15414.9) +- (0.00710904, 504.797)
};
\addlegendentry{mean, 320 units}
\addplot[solid]
coordinates {
(0.154797, 194468.0) +- (0.00706633, 14729.1)
(0.261136, 204072.0) +- (0.00852862, 14873.4)
(0.308136, 165186.0) +- (0.00773671, 10868.5)
(0.338818, 136626.0) +- (0.00808019, 7166.79)
(0.396532, 93971.1) +- (0.00752232, 4033.42)
(0.419071, 77841.0) +- (0.00731279, 3150.9)
(0.442737, 62844.3) +- (0.00752019, 2346.68)
(0.467174, 50721.5) +- (0.00753052, 1810.67)
(0.481076, 44449.1) +- (0.00752028, 1519.44)
};
\addlegendentry{mean, 640 units}
\addplot[thick, dotted]
coordinates {
(0.143942, 173514.0) +- (0.00910836, 0)
(0.206649, 262809.0) +- (0.00836592, 0)
(0.246022, 274765.0) +- (0.00812366, 0)
(0.271731, 326285.0) +- (0.00811458, 0)
(0.336609, 282708.0) +- (0.00746947, 0)
(0.355467, 500865.0) +- (0.00720303, 0)
(0.379126, 472882.0) +- (0.00751296, 0)
(0.400608, 427795.0) +- (0.00661102, 0)
(0.415544, 454066.0) +- (0.00650509, 0)
};
\addlegendentry{maximum, 160 units}
\addplot[thick, dashed]
coordinates {
(0.145098, 665884.0) +- (0.00629342, 0)
(0.238705, 982300.0) +- (0.00823697, 0)
(0.284467, 872547.0) +- (0.00814058, 0)
(0.310308, 899577.0) +- (0.00780967, 0)
(0.36885, 1103040.0) +- (0.00747156, 0)
(0.389151, 1241950.0) +- (0.00736526, 0)
(0.413018, 1045520.0) +- (0.00704032, 0)
(0.43612, 1135860.0) +- (0.00729952, 0)
(0.452103, 876445.0) +- (0.00710904, 0)
};
\addlegendentry{maximum, 320 units}
\addplot[thick, solid]
coordinates {
(0.154797, 2180610.0) +- (0.00706633, 0)
(0.261136, 3523880.0) +- (0.00852862, 0)
(0.308136, 4664960.0) +- (0.00773671, 0)
(0.338818, 2939580.0) +- (0.00808019, 0)
(0.396532, 4403590.0) +- (0.00752232, 0)
(0.419071, 5185010.0) +- (0.00731279, 0)
(0.442737, 4051880.0) +- (0.00752019, 0)
(0.467174, 5068690.0) +- (0.00753052, 0)
(0.481076, 5420570.0) +- (0.00752028, 0)
};
\addlegendentry{maximum, 640 units}
\end{semilogyaxis}
\end{tikzpicture}

%% file: Definitions/memory_10p_25.tex
\begin{tikzpicture}
\begin{semilogyaxis}[xlabel = {network utilization}, ylabel = {memory words}, height = 6.5 cm, width = 8.5 cm, grid = major, ticks = major, legend columns = 3, legend to name = regular, legend style = {/tikz/every even column/.append style = {column sep = 0.25 cm}}, xmin = 0.075, xmax = 0.4, ymin = 1e3, ymax = 1e8]
\addplot[dotted]
coordinates {
(0.113935, 8922.89) +- (0.00713754, 529.27)
(0.181142, 9200.2) +- (0.00732259, 503.835)
(0.225823, 8561.49) +- (0.00806008, 544.028)
(0.256669, 7460.27) +- (0.00708683, 409.374)
(0.305974, 5551.34) +- (0.00790537, 269.632)
(0.326675, 4860.62) +- (0.00689427, 198.597)
(0.353531, 3912.04) +- (0.00668977, 145.917)
(0.375944, 3233.75) +- (0.00702454, 112.651)
(0.390798, 2772.2) +- (0.00688805, 92.3923)
};
\addlegendentry{mean, 160 units}
\addplot[dashed]
coordinates {
(0.103148, 8384.18) +- (0.00672575, 490.781)
(0.180577, 9000.79) +- (0.00648534, 525.941)
(0.224838, 7791.43) +- (0.00759845, 402.263)
(0.254381, 7318.82) +- (0.00727339, 428.985)
(0.311084, 5111.33) +- (0.0076647, 248.288)
(0.328403, 4378.69) +- (0.00781389, 161.399)
(0.354752, 3590.06) +- (0.00712736, 134.033)
(0.375796, 2935.7) +- (0.00732195, 91.1981)
(0.391025, 2572.2) +- (0.00665829, 90.2506)
};
\addlegendentry{mean, 320 units}
\addplot[solid]
coordinates {
(0.107152, 8546.99) +- (0.00597721, 437.694)
(0.177023, 8863.48) +- (0.00679407, 481.761)
(0.22269, 7925.74) +- (0.00759882, 422.024)
(0.257486, 6947.73) +- (0.00785466, 389.386)
(0.311219, 4954.06) +- (0.0074807, 235.132)
(0.331221, 4338.23) +- (0.00729832, 188.646)
(0.354567, 3512.13) +- (0.00707648, 135.43)
(0.376776, 2795.57) +- (0.00730113, 92.2014)
(0.391779, 2412.63) +- (0.00706304, 74.1247)
};
\addlegendentry{mean, 640 units}
\addplot[thick, dotted]
coordinates {
(0.113935, 88258.0) +- (0.00713754, 0)
(0.181142, 113813.0) +- (0.00732259, 0)
(0.225823, 192566.0) +- (0.00806008, 0)
(0.256669, 114132.0) +- (0.00708683, 0)
(0.305974, 158063.0) +- (0.00790537, 0)
(0.326675, 190623.0) +- (0.00689427, 0)
(0.353531, 93805.0) +- (0.00668977, 0)
(0.375944, 100869.0) +- (0.00702454, 0)
(0.390798, 114344.0) +- (0.00688805, 0)
};
\addlegendentry{maximum, 160 units}
\addplot[thick, dashed]
coordinates {
(0.103148, 65668.0) +- (0.00672575, 0)
(0.180577, 102943.0) +- (0.00648534, 0)
(0.224838, 116731.0) +- (0.00759845, 0)
(0.254381, 153047.0) +- (0.00727339, 0)
(0.311084, 150673.0) +- (0.0076647, 0)
(0.328403, 84151.0) +- (0.00781389, 0)
(0.354752, 73758.0) +- (0.00712736, 0)
(0.375796, 68705.0) +- (0.00732195, 0)
(0.391025, 79550.0) +- (0.00665829, 0)
};
\addlegendentry{maximum, 320 units}
\addplot[thick, solid]
coordinates {
(0.107152, 80364.0) +- (0.00597721, 0)
(0.177023, 92839.0) +- (0.00679407, 0)
(0.22269, 88922.0) +- (0.00759882, 0)
(0.257486, 140898.0) +- (0.00785466, 0)
(0.311219, 108626.0) +- (0.0074807, 0)
(0.331221, 79476.0) +- (0.00729832, 0)
(0.354567, 69297.0) +- (0.00707648, 0)
(0.376776, 87299.0) +- (0.00730113, 0)
(0.391779, 47237.0) +- (0.00706304, 0)
};
\addlegendentry{maximum, 640 units}
\end{semilogyaxis}
\end{tikzpicture}

%% file: Definitions/memory_10_50.tex
\begin{tikzpicture}
\begin{semilogyaxis}[xlabel = {network utilization}, ylabel = {memory words}, height = 6.5 cm, width = 8.5 cm, grid = major, ticks = major, legend columns = 3, legend to name = regular, legend style = {/tikz/every even column/.append style = {column sep = 0.25 cm}}, xmin = 0.1, xmax = 0.5, ymin = 1e3, ymax = 1e10]
\addplot[dotted]
coordinates {
(0.137347, 123801.0) +- (0.00417829, 6831.98)
(0.209375, 128726.0) +- (0.00506164, 7607.14)
(0.246112, 111611.0) +- (0.00501929, 6251.49)
(0.268306, 102993.0) +- (0.00549038, 5878.98)
(0.314408, 69903.1) +- (0.0049428, 3020.51)
(0.334647, 55533.3) +- (0.00497782, 2466.73)
(0.351075, 47949.8) +- (0.00485923, 2075.36)
(0.370752, 36850.4) +- (0.00458584, 1343.83)
(0.386705, 30787.9) +- (0.00466367, 1095.46)
};
\addlegendentry{mean, 160 units}
\addplot[dashed]
coordinates {
(0.143737, 477719.0) +- (0.00423325, 29066.4)
(0.239671, 444603.0) +- (0.00541046, 30736.7)
(0.276411, 362597.0) +- (0.00502445, 21595.6)
(0.301211, 298764.0) +- (0.00473804, 16052.7)
(0.345596, 192778.0) +- (0.00477823, 9243.94)
(0.362728, 155443.0) +- (0.00498939, 6063.0)
(0.383287, 122449.0) +- (0.00486709, 4767.31)
(0.403796, 92212.5) +- (0.0046782, 3199.07)
(0.418844, 77791.8) +- (0.0045363, 2671.93)
};
\addlegendentry{mean, 320 units}
\addplot[solid]
coordinates {
(0.156029, 1970710.0) +- (0.00508507, 136064.0)
(0.245737, 1642880.0) +- (0.00482973, 126290.0)
(0.296806, 1129120.0) +- (0.00492433, 71059.6)
(0.321653, 946066.0) +- (0.00455477, 57656.2)
(0.36933, 556457.0) +- (0.00514214, 26056.1)
(0.387559, 438384.0) +- (0.00519268, 18372.8)
(0.409804, 337065.0) +- (0.0048596, 13637.0)
(0.428943, 265922.0) +- (0.00481172, 9933.15)
(0.443752, 223045.0) +- (0.00508116, 7475.13)
};
\addlegendentry{mean, 640 units}
\addplot[thick, dotted]
coordinates {
(0.137347, 1947550.0) +- (0.00417829, 0)
(0.209375, 2398510.0) +- (0.00506164, 0)
(0.246112, 1956790.0) +- (0.00501929, 0)
(0.268306, 2810960.0) +- (0.00549038, 0)
(0.314408, 2764940.0) +- (0.0049428, 0)
(0.334647, 3618620.0) +- (0.00497782, 0)
(0.351075, 6671040.0) +- (0.00485923, 0)
(0.370752, 3425060.0) +- (0.00458584, 0)
(0.386705, 4721150.0) +- (0.00466367, 0)
};
\addlegendentry{maximum, 160 units}
\addplot[thick, dashed]
coordinates {
(0.143737, 7522520.0) +- (0.00423325, 0)
(0.239671, 13573600.0) +- (0.00541046, 0)
(0.276411, 12827700.0) +- (0.00502445, 0)
(0.301211, 11348100.0) +- (0.00473804, 0)
(0.345596, 21966500.0) +- (0.00477823, 0)
(0.362728, 13905700.0) +- (0.00498939, 0)
(0.383287, 15480300.0) +- (0.00486709, 0)
(0.403796, 12795600.0) +- (0.0046782, 0)
(0.418844, 14288100.0) +- (0.0045363, 0)
};
\addlegendentry{maximum, 320 units}
\addplot[thick, solid]
coordinates {
(0.156029, 31486500.0) +- (0.00508507, 0)
(0.245737, 40132200.0) +- (0.00482973, 0)
(0.296806, 42288200.0) +- (0.00492433, 0)
(0.321653, 44845800.0) +- (0.00455477, 0)
(0.36933, 47388500.0) +- (0.00514214, 0)
(0.387559, 48288700.0) +- (0.00519268, 0)
(0.409804, 56500700.0) +- (0.0048596, 0)
(0.428943, 65844000.0) +- (0.00481172, 0)
(0.443752, 56016900.0) +- (0.00508116, 0)
};
\addlegendentry{maximum, 640 units}
\end{semilogyaxis}
\end{tikzpicture}

%% file: Definitions/memory_10p_50.tex
\begin{tikzpicture}
\begin{semilogyaxis}[xlabel = {network utilization}, ylabel = {memory words}, height = 6.5 cm, width = 8.5 cm, grid = major, ticks = major, legend columns = 3, legend to name = regular, legend style = {/tikz/every even column/.append style = {column sep = 0.25 cm}}, xmin = 0.075, xmax = 0.4, ymin = 1e3, ymax = 1e8]
\addplot[dotted]
coordinates {
(0.117661, 62720.9) +- (0.00569911, 3599.28)
(0.188487, 61350.3) +- (0.00599344, 3480.58)
(0.220736, 55927.6) +- (0.0059305, 3094.05)
(0.247075, 48509.7) +- (0.00565563, 2356.94)
(0.291796, 33754.9) +- (0.00524972, 1556.2)
(0.31307, 27457.9) +- (0.00579297, 1129.17)
(0.332693, 21015.2) +- (0.00505063, 754.687)
(0.351376, 16637.3) +- (0.00446423, 623.696)
(0.362726, 13953.8) +- (0.00483861, 440.895)
};
\addlegendentry{mean, 160 units}
\addplot[dashed]
coordinates {
(0.102604, 54852.7) +- (0.00513301, 3337.28)
(0.181763, 57566.5) +- (0.00564969, 2922.6)
(0.226664, 51441.3) +- (0.00565136, 2600.88)
(0.244222, 46488.7) +- (0.00534454, 2584.78)
(0.297387, 29552.0) +- (0.004939, 1273.48)
(0.31379, 23838.0) +- (0.00565221, 912.243)
(0.334351, 18709.1) +- (0.00489876, 711.614)
(0.350942, 14701.6) +- (0.00452578, 497.683)
(0.364262, 12203.1) +- (0.00494357, 372.351)
};
\addlegendentry{mean, 320 units}
\addplot[solid]
coordinates {
(0.0966696, 49678.2) +- (0.00420626, 2619.17)
(0.188585, 56780.1) +- (0.00615896, 3073.58)
(0.228492, 50410.3) +- (0.005742, 2698.39)
(0.249141, 43296.9) +- (0.00559674, 2435.92)
(0.293901, 29179.2) +- (0.00538317, 1308.32)
(0.3112, 23835.7) +- (0.00549271, 935.775)
(0.331432, 18065.2) +- (0.00564668, 674.233)
(0.349545, 13998.5) +- (0.00518425, 508.803)
(0.364471, 11300.9) +- (0.00502601, 365.299)
};
\addlegendentry{mean, 640 units}
\addplot[thick, dotted]
coordinates {
(0.117661, 753462.0) +- (0.00569911, 0)
(0.188487, 951997.0) +- (0.00599344, 0)
(0.220736, 1714440.0) +- (0.0059305, 0)
(0.247075, 1216800.0) +- (0.00565563, 0)
(0.291796, 972768.0) +- (0.00524972, 0)
(0.31307, 1132920.0) +- (0.00579297, 0)
(0.332693, 1241940.0) +- (0.00505063, 0)
(0.351376, 1093870.0) +- (0.00446423, 0)
(0.362726, 912570.0) +- (0.00483861, 0)
};
\addlegendentry{maximum, 160 units}
\addplot[thick, dashed]
coordinates {
(0.102604, 583906.0) +- (0.00513301, 0)
(0.181763, 683657.0) +- (0.00564969, 0)
(0.226664, 841370.0) +- (0.00565136, 0)
(0.244222, 1072730.0) +- (0.00534454, 0)
(0.297387, 1183730.0) +- (0.004939, 0)
(0.31379, 799286.0) +- (0.00565221, 0)
(0.334351, 808417.0) +- (0.00489876, 0)
(0.350942, 598418.0) +- (0.00452578, 0)
(0.364262, 647586.0) +- (0.00494357, 0)
};
\addlegendentry{maximum, 320 units}
\addplot[thick, solid]
coordinates {
(0.0966696, 611689.0) +- (0.00420626, 0)
(0.188585, 764676.0) +- (0.00615896, 0)
(0.228492, 808053.0) +- (0.005742, 0)
(0.249141, 933236.0) +- (0.00559674, 0)
(0.293901, 655783.0) +- (0.00538317, 0)
(0.3112, 936511.0) +- (0.00549271, 0)
(0.331432, 657171.0) +- (0.00564668, 0)
(0.349545, 367564.0) +- (0.00518425, 0)
(0.364471, 352176.0) +- (0.00502601, 0)
};
\addlegendentry{maximum, 640 units}
\end{semilogyaxis}
\end{tikzpicture}

%% file: Definitions/memory_10_100.tex
\begin{tikzpicture}
\begin{semilogyaxis}[xlabel = {network utilization}, ylabel = {memory words}, height = 6.5 cm, width = 8.5 cm, grid = major, ticks = major, legend columns = 3, legend to name = regular, legend style = {/tikz/every even column/.append style = {column sep = 0.25 cm}}, xmin = 0.1, xmax = 0.5, ymin = 1e3, ymax = 1e10]
\addplot[dotted]
coordinates {
(0.138656, 1014610.0) +- (0.00349105, 55256.7)
(0.217396, 968050.0) +- (0.00338113, 48661.3)
(0.250139, 834224.0) +- (0.00342693, 38507.1)
(0.265751, 724013.0) +- (0.00358388, 31111.5)
(0.304259, 437297.0) +- (0.00306796, 17666.4)
(0.314785, 379891.0) +- (0.00331254, 13834.9)
(0.333959, 280020.0) +- (0.00298775, 8877.79)
(0.350223, 213947.0) +- (0.00274093, 6662.14)
(0.361814, 169221.0) +- (0.00292915, 4831.18)
};
\addlegendentry{mean, 160 units}
\addplot[dashed]
coordinates {
(0.158062, 4401170.0) +- (0.00276868, 236643.0)
(0.2369, 3437930.0) +- (0.00294746, 181403.0)
(0.271127, 2552310.0) +- (0.00297365, 126020.0)
(0.291183, 2022020.0) +- (0.00290559, 92506.1)
(0.327215, 1239090.0) +- (0.00306662, 46316.4)
(0.341482, 989102.0) +- (0.0030209, 35768.0)
(0.358734, 717942.0) +- (0.00307359, 23112.3)
(0.375301, 552642.0) +- (0.00269418, 16625.8)
(0.386229, 453086.0) +- (0.00275794, 12868.3)
};
\addlegendentry{mean, 320 units}
\addplot[solid]
coordinates {
(0.148295, 19056000.0) +- (0.00295821, 1025070.0)
(0.238104, 12561400.0) +- (0.00267535, 818242.0)
(0.284313, 8629700.0) +- (0.00252097, 478305.0)
(0.307893, 6403010.0) +- (0.00301953, 295306.0)
(0.347749, 3455310.0) +- (0.00321284, 126248.0)
(0.363006, 2724080.0) +- (0.00312716, 96612.2)
(0.380794, 2028640.0) +- (0.00298298, 65936.8)
(0.397339, 1545250.0) +- (0.0029367, 48187.3)
(0.409587, 1247690.0) +- (0.00287135, 34611.3)
};
\addlegendentry{mean, 640 units}
\addplot[thick, dotted]
coordinates {
(0.138656, 23517200.0) +- (0.00349105, 0)
(0.217396, 41250000.0) +- (0.00338113, 0)
(0.250139, 26540500.0) +- (0.00342693, 0)
(0.265751, 27950900.0) +- (0.00358388, 0)
(0.304259, 61281500.0) +- (0.00306796, 0)
(0.314785, 54434100.0) +- (0.00331254, 0)
(0.333959, 36196200.0) +- (0.00298775, 0)
(0.350223, 36142200.0) +- (0.00274093, 0)
(0.361814, 32055400.0) +- (0.00292915, 0)
};
\addlegendentry{maximum, 160 units}
\addplot[thick, dashed]
coordinates {
(0.158062, 76278400.0) +- (0.00276868, 0)
(0.2369, 154759000.0) +- (0.00294746, 0)
(0.271127, 107775000.0) +- (0.00297365, 0)
(0.291183, 116663000.0) +- (0.00290559, 0)
(0.327215, 113403000.0) +- (0.00306662, 0)
(0.341482, 126368000.0) +- (0.0030209, 0)
(0.358734, 118878000.0) +- (0.00307359, 0)
(0.375301, 129484000.0) +- (0.00269418, 0)
(0.386229, 111008000.0) +- (0.00275794, 0)
};
\addlegendentry{maximum, 320 units}
\addplot[thick, solid]
coordinates {
(0.148295, 380703000.0) +- (0.00295821, 0)
(0.238104, 539637000.0) +- (0.00267535, 0)
(0.284313, 632053000.0) +- (0.00252097, 0)
(0.307893, 491030000.0) +- (0.00301953, 0)
(0.347749, 305250000.0) +- (0.00321284, 0)
(0.363006, 497552000.0) +- (0.00312716, 0)
(0.380794, 517948000.0) +- (0.00298298, 0)
(0.397339, 1168270000.0) +- (0.0029367, 0)
(0.409587, 837635000.0) +- (0.00287135, 0)
};
\addlegendentry{maximum, 640 units}
\end{semilogyaxis}
\end{tikzpicture}

%% file: Definitions/memory_10p_100.tex
\begin{tikzpicture}
\begin{semilogyaxis}[xlabel = {network utilization}, ylabel = {memory words}, height = 6.5 cm, width = 8.5 cm, grid = major, ticks = major, legend columns = 3, legend to name = regular, legend style = {/tikz/every even column/.append style = {column sep = 0.25 cm}}, xmin = 0.075, xmax = 0.4, ymin = 1e3, ymax = 1e8]
\addplot[dotted]
coordinates {
(0.119977, 428514.0) +- (0.00348355, 20384.1)
(0.190231, 429419.0) +- (0.0039338, 21052.6)
(0.226321, 370603.0) +- (0.00411252, 16973.5)
(0.243865, 330418.0) +- (0.00379688, 14019.3)
(0.288175, 202775.0) +- (0.00338089, 7663.16)
(0.299012, 169170.0) +- (0.00349773, 5954.04)
(0.316203, 125278.0) +- (0.00346146, 4215.89)
(0.329426, 95625.5) +- (0.00298566, 2872.81)
(0.341588, 76639.6) +- (0.00293003, 2063.47)
};
\addlegendentry{mean, 160 units}
\addplot[dashed]
coordinates {
(0.11443, 394657.0) +- (0.00335236, 17075.4)
(0.188332, 413845.0) +- (0.0038314, 17394.9)
(0.226132, 350976.0) +- (0.00412051, 16004.2)
(0.248117, 300322.0) +- (0.00396907, 15168.4)
(0.288263, 185024.0) +- (0.00365752, 7438.69)
(0.301603, 143831.0) +- (0.0034388, 5275.81)
(0.315909, 109027.0) +- (0.00315253, 3544.22)
(0.330948, 82390.0) +- (0.00306049, 2511.94)
(0.342636, 65800.2) +- (0.00289001, 1773.47)
};
\addlegendentry{mean, 320 units}
\addplot[solid]
coordinates {
(0.103775, 397484.0) +- (0.00298092, 17386.0)
(0.197487, 413623.0) +- (0.00356882, 20540.0)
(0.231169, 329044.0) +- (0.00336791, 16687.9)
(0.247417, 287927.0) +- (0.0037526, 12062.2)
(0.288076, 174272.0) +- (0.00345113, 7167.99)
(0.303308, 135695.0) +- (0.00356083, 4440.52)
(0.31731, 100845.0) +- (0.00361482, 3470.82)
(0.331111, 74932.3) +- (0.00330166, 2145.18)
(0.341973, 60967.3) +- (0.00287774, 1797.06)
};
\addlegendentry{mean, 640 units}
\addplot[thick, dotted]
coordinates {
(0.119977, 6875830.0) +- (0.00348355, 0)
(0.190231, 10374300.0) +- (0.0039338, 0)
(0.226321, 10152300.0) +- (0.00411252, 0)
(0.243865, 9347060.0) +- (0.00379688, 0)
(0.288175, 14183800.0) +- (0.00338089, 0)
(0.299012, 11254700.0) +- (0.00349773, 0)
(0.316203, 11666600.0) +- (0.00346146, 0)
(0.329426, 7628080.0) +- (0.00298566, 0)
(0.341588, 7167850.0) +- (0.00293003, 0)
};
\addlegendentry{maximum, 160 units}
\addplot[thick, dashed]
coordinates {
(0.11443, 6265260.0) +- (0.00335236, 0)
(0.188332, 7473470.0) +- (0.0038314, 0)
(0.226132, 6893170.0) +- (0.00412051, 0)
(0.248117, 9815350.0) +- (0.00396907, 0)
(0.288263, 8173050.0) +- (0.00365752, 0)
(0.301603, 7284690.0) +- (0.0034388, 0)
(0.315909, 4960540.0) +- (0.00315253, 0)
(0.330948, 5132850.0) +- (0.00306049, 0)
(0.342636, 5789400.0) +- (0.00289001, 0)
};
\addlegendentry{maximum, 320 units}
\addplot[thick, solid]
coordinates {
(0.103775, 5439470.0) +- (0.00298092, 0)
(0.197487, 10035500.0) +- (0.00356882, 0)
(0.231169, 4682990.0) +- (0.00336791, 0)
(0.247417, 6913890.0) +- (0.0037526, 0)
(0.288076, 8174170.0) +- (0.00345113, 0)
(0.303308, 7178470.0) +- (0.00356083, 0)
(0.31731, 3499740.0) +- (0.00361482, 0)
(0.331111, 2880200.0) +- (0.00330166, 0)
(0.341973, 2879630.0) +- (0.00287774, 0)
};
\addlegendentry{maximum, 640 units}
\end{semilogyaxis}
\end{tikzpicture}

%% file: Definitions/stack_25.tex
\begin{tikzpicture}
\begin{axis}[xlabel = {network utilization}, ylabel = {memory words}, stack plots = y, area style, enlarge x limits = false, height = 3.6 cm, width = 5.6 cm, ymin = 0, legend columns = -1, legend to name = stack, legend style = {/tikz/every even column/.append style = {column sep = 0.25 cm}}, ]
\addplot[pattern = north east lines]
coordinates { (0.145098, 664335.0) +- (0.00629342, 0)
(0.238705, 979080.0) +- (0.00823697, 0)
(0.284467, 867675.0) +- (0.00814058, 0)
(0.310308, 898215.0) +- (0.00780967, 0)
(0.36885, 1035615.0) +- (0.00747156, 0)
(0.389151, 1108740.0) +- (0.00736526, 0)
(0.413018, 1043640.0) +- (0.00704032, 0)
(0.43612, 1067250.0) +- (0.00729952, 0)
(0.452103, 808560.0) +- (0.00710904, 0)
}\closedcycle;
\addlegendentry{permanent labels}
\addplot[pattern = crosshatch]
coordinates { (0.145098, 130245.0) +- (0.00629342, 0)
(0.238705, 141405.0) +- (0.00823697, 0)
(0.284467, 156075.0) +- (0.00814058, 0)
(0.310308, 183780.0) +- (0.00780967, 0)
(0.36885, 221820.0) +- (0.00747156, 0)
(0.389151, 216705.0) +- (0.00736526, 0)
(0.413018, 291450.0) +- (0.00704032, 0)
(0.43612, 255750.0) +- (0.00729952, 0)
(0.452103, 165120.0) +- (0.00710904, 0)
}\closedcycle;
\addlegendentry{tentative labels}
\addplot[pattern = north west lines]
coordinates { (0.145098, 18252.0) +- (0.00629342, 0)
(0.238705, 19876.0) +- (0.00823697, 0)
(0.284467, 21932.0) +- (0.00814058, 0)
(0.310308, 25522.0) +- (0.00780967, 0)
(0.36885, 30404.0) +- (0.00747156, 0)
(0.389151, 29284.0) +- (0.00736526, 0)
(0.413018, 39810.0) +- (0.00704032, 0)
(0.43612, 35314.0) +- (0.00729952, 0)
(0.452103, 22488.0) +- (0.00710904, 0)
}\closedcycle;
\addlegendentry{priority queue elements}
\end{axis}
\end{tikzpicture}

%% file: Definitions/stack_50.tex
\begin{tikzpicture}
\begin{axis}[xlabel = {network utilization}, ylabel = {memory words}, stack plots = y, area style, enlarge x limits = false, height = 3.6 cm, width = 5.6 cm, ymin = 0, legend columns = -1, legend to name = stack, legend style = {/tikz/every even column/.append style = {column sep = 0.25 cm}}, ]
\addplot[pattern = north east lines]
coordinates { (0.143737, 7504155.0) +- (0.00423325, 0)
(0.239671, 13566780.0) +- (0.00541046, 0)
(0.276411, 12808995.0) +- (0.00502445, 0)
(0.301211, 11341380.0) +- (0.00473804, 0)
(0.345596, 21921900.0) +- (0.00477823, 0)
(0.362728, 13898880.0) +- (0.00498939, 0)
(0.383287, 15478650.0) +- (0.00486709, 0)
(0.403796, 12795420.0) +- (0.0046782, 0)
(0.418844, 14286690.0) +- (0.0045363, 0)
}\closedcycle;
\addlegendentry{permanent labels}
\addplot[pattern = crosshatch]
coordinates { (0.143737, 1315395.0) +- (0.00423325, 0)
(0.239671, 1515270.0) +- (0.00541046, 0)
(0.276411, 1373730.0) +- (0.00502445, 0)
(0.301211, 1637550.0) +- (0.00473804, 0)
(0.345596, 2333715.0) +- (0.00477823, 0)
(0.362728, 3113025.0) +- (0.00498939, 0)
(0.383287, 1815345.0) +- (0.00486709, 0)
(0.403796, 1864485.0) +- (0.0046782, 0)
(0.418844, 2049945.0) +- (0.0045363, 0)
}\closedcycle;
\addlegendentry{tentative labels}
\addplot[pattern = north west lines]
coordinates { (0.143737, 180458.0) +- (0.00423325, 0)
(0.239671, 210676.0) +- (0.00541046, 0)
(0.276411, 189642.0) +- (0.00502445, 0)
(0.301211, 223518.0) +- (0.00473804, 0)
(0.345596, 318926.0) +- (0.00477823, 0)
(0.362728, 426324.0) +- (0.00498939, 0)
(0.383287, 249640.0) +- (0.00486709, 0)
(0.403796, 254886.0) +- (0.0046782, 0)
(0.418844, 277632.0) +- (0.0045363, 0)
}\closedcycle;
\addlegendentry{priority queue elements}
\end{axis}
\end{tikzpicture}

%% file: Definitions/stack_100.tex
\begin{tikzpicture}
\begin{axis}[xlabel = {network utilization}, ylabel = {memory words}, stack plots = y, area style, enlarge x limits = false, height = 3.6 cm, width = 5.6 cm, ymin = 0, legend columns = -1, legend to name = stack, legend style = {/tikz/every even column/.append style = {column sep = 0.25 cm}}, ]
\addplot[pattern = north east lines]
coordinates { (0.158062, 74322150.0) +- (0.00276868, 0)
(0.2369, 154726500.0) +- (0.00294746, 0)
(0.271127, 93805650.0) +- (0.00297365, 0)
(0.291183, 116647350.0) +- (0.00290559, 0)
(0.327215, 113369100.0) +- (0.00306662, 0)
(0.341482, 116007000.0) +- (0.0030209, 0)
(0.358734, 117113100.0) +- (0.00307359, 0)
(0.375301, 129468300.0) +- (0.00269418, 0)
(0.386229, 110976900.0) +- (0.00275794, 0)
}\closedcycle;
\addlegendentry{permanent labels}
\addplot[pattern = crosshatch]
coordinates { (0.158062, 9864825.0) +- (0.00276868, 0)
(0.2369, 10964610.0) +- (0.00294746, 0)
(0.271127, 21431400.0) +- (0.00297365, 0)
(0.291183, 15501750.0) +- (0.00290559, 0)
(0.327215, 15412050.0) +- (0.00306662, 0)
(0.341482, 16620150.0) +- (0.0030209, 0)
(0.358734, 12778815.0) +- (0.00307359, 0)
(0.375301, 12649785.0) +- (0.00269418, 0)
(0.386229, 18386700.0) +- (0.00275794, 0)
}\closedcycle;
\addlegendentry{tentative labels}
\addplot[pattern = north west lines]
coordinates { (0.158062, 1431838.0) +- (0.00276868, 0)
(0.2369, 1511214.0) +- (0.00294746, 0)
(0.271127, 2985000.0) +- (0.00297365, 0)
(0.291183, 2152640.0) +- (0.00290559, 0)
(0.327215, 2139000.0) +- (0.00306662, 0)
(0.341482, 2295380.0) +- (0.0030209, 0)
(0.358734, 1754308.0) +- (0.00307359, 0)
(0.375301, 1749266.0) +- (0.00269418, 0)
(0.386229, 2505980.0) +- (0.00275794, 0)
}\closedcycle;
\addlegendentry{priority queue elements}
\end{axis}
\end{tikzpicture}

%% file: Definitions/bbp_10_25.tex
\begin{tikzpicture}
\begin{semilogyaxis}[xlabel = {network utilization}, ylabel = {bandwidth block-\\ing probability}, height = 6.5 cm, width = 8 cm, grid = both, ylabel style = {align = center, xshift = -0.2 cm}, legend columns = 3, legend to name = bbp, legend style = {/tikz/every even column/.append style = {column sep = 0.25 cm}}, xmin = 0.1, xmax = 0.5, ymin = 0.1, ymax = 1, ]
\addplot[dotted]
coordinates {
(0.143942, 0.206205) +- (0.00910836, 0.0179962)
(0.206649, 0.327936) +- (0.00836592, 0.0171455)
(0.246022, 0.427627) +- (0.00812366, 0.0142697)
(0.271731, 0.509851) +- (0.00811458, 0.0119816)
(0.336609, 0.679086) +- (0.00746947, 0.00770858)
(0.355467, 0.73487) +- (0.00720303, 0.00715409)
(0.379126, 0.792704) +- (0.00751296, 0.0050537)
(0.400608, 0.832996) +- (0.00661102, 0.00440617)
(0.415544, 0.855794) +- (0.00650509, 0.00374163)
};
\addlegendentry{proposed algorithm, 160 units}
\addplot[dashed]
coordinates {
(0.145098, 0.154141) +- (0.00629342, 0.0205183)
(0.238705, 0.28112) +- (0.00823697, 0.017828)
(0.284467, 0.383399) +- (0.00814058, 0.0159161)
(0.310308, 0.448743) +- (0.00780967, 0.0137347)
(0.36885, 0.638466) +- (0.00747156, 0.00824032)
(0.389151, 0.704616) +- (0.00736526, 0.00719728)
(0.413018, 0.764638) +- (0.00704032, 0.00571748)
(0.43612, 0.810201) +- (0.00729952, 0.00482058)
(0.452103, 0.838693) +- (0.00710904, 0.00414951)
};
\addlegendentry{proposed algorithm, 320 units}
\addplot[solid]
coordinates {
(0.154797, 0.13511) +- (0.00706633, 0.0229225)
(0.261136, 0.222572) +- (0.00852862, 0.0192651)
(0.308136, 0.314722) +- (0.00773671, 0.0175187)
(0.338818, 0.399804) +- (0.00808019, 0.0142139)
(0.396532, 0.6039) +- (0.00752232, 0.00900276)
(0.419071, 0.67668) +- (0.00731279, 0.00760664)
(0.442737, 0.742477) +- (0.00752019, 0.00614124)
(0.467174, 0.795918) +- (0.00753052, 0.00487963)
(0.481076, 0.825662) +- (0.00752028, 0.00429553)
};
\addlegendentry{proposed algorithm, 640 units}
\addplot[thick, dotted]
coordinates {
(0.143942, 0.224674) +- (0.00910836, 0.0186198)
(0.206649, 0.347268) +- (0.00836592, 0.0171067)
(0.246022, 0.452565) +- (0.00812366, 0.0139056)
(0.271731, 0.532525) +- (0.00811458, 0.011493)
(0.336609, 0.69709) +- (0.00746947, 0.00733704)
(0.355467, 0.748319) +- (0.00720303, 0.00679658)
(0.379126, 0.801388) +- (0.00751296, 0.00483001)
(0.400608, 0.839151) +- (0.00661102, 0.0042675)
(0.415544, 0.860614) +- (0.00650509, 0.0036102)
};
\addlegendentry{edge-exclusion, 160 units}
\addplot[thick, dashed]
coordinates {
(0.145098, 0.168194) +- (0.00629342, 0.0210589)
(0.238705, 0.307283) +- (0.00823697, 0.0175839)
(0.284467, 0.414234) +- (0.00814058, 0.0149569)
(0.310308, 0.478425) +- (0.00780967, 0.0132995)
(0.36885, 0.658986) +- (0.00747156, 0.00778043)
(0.389151, 0.720499) +- (0.00736526, 0.00674192)
(0.413018, 0.776048) +- (0.00704032, 0.00548284)
(0.43612, 0.817781) +- (0.00729952, 0.00460979)
(0.452103, 0.843994) +- (0.00710904, 0.00399587)
};
\addlegendentry{edge-exclusion, 320 units}
\addplot[thick, solid]
coordinates {
(0.154797, 0.147021) +- (0.00706633, 0.0231021)
(0.261136, 0.249879) +- (0.00852862, 0.0188476)
(0.308136, 0.345962) +- (0.00773671, 0.016996)
(0.338818, 0.432424) +- (0.00808019, 0.013648)
(0.396532, 0.62708) +- (0.00752232, 0.00863616)
(0.419071, 0.693987) +- (0.00731279, 0.00702735)
(0.442737, 0.754971) +- (0.00752019, 0.00583208)
(0.467174, 0.804368) +- (0.00753052, 0.00461781)
(0.481076, 0.83167) +- (0.00752028, 0.00411038)
};
\addlegendentry{edge-exclusion, 640 units}
\end{semilogyaxis}
\end{tikzpicture}

%% file: Definitions/bbp_10p_25.tex
\begin{tikzpicture}
\begin{semilogyaxis}[xlabel = {network utilization}, ylabel = {bandwidth block-\\ing probability}, height = 6.5 cm, width = 8 cm, grid = both, ylabel style = {align = center, xshift = -0.2 cm}, legend columns = 3, legend to name = bbp, legend style = {/tikz/every even column/.append style = {column sep = 0.25 cm}}, xmin = 0.09, xmax = 0.4, ymin = 0.1, ymax = 1, ]
\addplot[dotted]
coordinates {
(0.113935, 0.228494) +- (0.00713754, 0.0246204)
(0.181142, 0.376098) +- (0.00732259, 0.0184097)
(0.225823, 0.486652) +- (0.00806008, 0.0167387)
(0.256669, 0.565136) +- (0.00708683, 0.0130923)
(0.305974, 0.70579) +- (0.00790537, 0.00818434)
(0.326675, 0.756791) +- (0.00689427, 0.00640885)
(0.353531, 0.811584) +- (0.00668977, 0.00504121)
(0.375944, 0.843662) +- (0.00702454, 0.00434553)
(0.390798, 0.864917) +- (0.00688805, 0.00380281)
};
\addlegendentry{proposed algorithm, 160 units}
\addplot[dashed]
coordinates {
(0.103148, 0.206517) +- (0.00672575, 0.0271048)
(0.180577, 0.34211) +- (0.00648534, 0.0229632)
(0.224838, 0.480086) +- (0.00759845, 0.0154151)
(0.254381, 0.555967) +- (0.00727339, 0.015515)
(0.311084, 0.702266) +- (0.0076647, 0.00767466)
(0.328403, 0.756082) +- (0.00781389, 0.00617099)
(0.354752, 0.805693) +- (0.00712736, 0.00519731)
(0.375796, 0.840853) +- (0.00732195, 0.00409484)
(0.391025, 0.86065) +- (0.00665829, 0.00361893)
};
\addlegendentry{proposed algorithm, 320 units}
\addplot[solid]
coordinates {
(0.107152, 0.206948) +- (0.00597721, 0.0224391)
(0.177023, 0.334423) +- (0.00679407, 0.018978)
(0.22269, 0.4632) +- (0.00759882, 0.01565)
(0.257486, 0.546273) +- (0.00785466, 0.0130223)
(0.311219, 0.697809) +- (0.0074807, 0.00839267)
(0.331221, 0.748929) +- (0.00729832, 0.0069053)
(0.354567, 0.802641) +- (0.00707648, 0.00530364)
(0.376776, 0.842158) +- (0.00730113, 0.00453058)
(0.391779, 0.861279) +- (0.00706304, 0.00374162)
};
\addlegendentry{proposed algorithm, 640 units}
\addplot[thick, dotted]
coordinates {
(0.113935, 0.244171) +- (0.00713754, 0.0248071)
(0.181142, 0.397861) +- (0.00732259, 0.0184817)
(0.225823, 0.505956) +- (0.00806008, 0.0169231)
(0.256669, 0.584551) +- (0.00708683, 0.0132697)
(0.305974, 0.720207) +- (0.00790537, 0.00776262)
(0.326675, 0.768196) +- (0.00689427, 0.0062991)
(0.353531, 0.820252) +- (0.00668977, 0.00472282)
(0.375944, 0.849399) +- (0.00702454, 0.00421634)
(0.390798, 0.86916) +- (0.00688805, 0.00373993)
};
\addlegendentry{edge-exclusion, 160 units}
\addplot[thick, dashed]
coordinates {
(0.103148, 0.22537) +- (0.00672575, 0.0276096)
(0.180577, 0.363878) +- (0.00648534, 0.0230556)
(0.224838, 0.5007) +- (0.00759845, 0.0157485)
(0.254381, 0.575359) +- (0.00727339, 0.0152698)
(0.311084, 0.717655) +- (0.0076647, 0.00716607)
(0.328403, 0.767872) +- (0.00781389, 0.00589808)
(0.354752, 0.813672) +- (0.00712736, 0.0049809)
(0.375796, 0.846445) +- (0.00732195, 0.0040458)
(0.391025, 0.864889) +- (0.00665829, 0.00359221)
};
\addlegendentry{edge-exclusion, 320 units}
\addplot[thick, solid]
coordinates {
(0.107152, 0.224145) +- (0.00597721, 0.0228235)
(0.177023, 0.358082) +- (0.00679407, 0.0184293)
(0.22269, 0.487494) +- (0.00759882, 0.015578)
(0.257486, 0.56681) +- (0.00785466, 0.0127661)
(0.311219, 0.711979) +- (0.0074807, 0.00840943)
(0.331221, 0.760621) +- (0.00729832, 0.00666802)
(0.354567, 0.810732) +- (0.00707648, 0.00507093)
(0.376776, 0.848102) +- (0.00730113, 0.00447166)
(0.391779, 0.865823) +- (0.00706304, 0.0036355)
};
\addlegendentry{edge-exclusion, 640 units}
\end{semilogyaxis}
\end{tikzpicture}

%% file: Definitions/bbp_10_50.tex
\begin{tikzpicture}
\begin{semilogyaxis}[xlabel = {network utilization}, ylabel = {bandwidth block-\\ing probability}, height = 6.5 cm, width = 8 cm, grid = both, ylabel style = {align = center, xshift = -0.2 cm}, legend columns = 3, legend to name = bbp, legend style = {/tikz/every even column/.append style = {column sep = 0.25 cm}}, xmin = 0.1, xmax = 0.5, ymin = 0.07, ymax = 1, ]
\addplot[dotted]
coordinates {
(0.137347, 0.17234) +- (0.00417829, 0.0138866)
(0.209375, 0.301411) +- (0.00506164, 0.0121947)
(0.246112, 0.421472) +- (0.00501929, 0.0115688)
(0.268306, 0.480677) +- (0.00549038, 0.0092737)
(0.314408, 0.649877) +- (0.0049428, 0.00611453)
(0.334647, 0.717617) +- (0.00497782, 0.00463977)
(0.351075, 0.769408) +- (0.00485923, 0.00426343)
(0.370752, 0.814429) +- (0.00458584, 0.00339042)
(0.386705, 0.838913) +- (0.00466367, 0.00285873)
};
\addlegendentry{proposed algorithm, 160 units}
\addplot[dashed]
coordinates {
(0.143737, 0.103505) +- (0.00423325, 0.0136787)
(0.239671, 0.231692) +- (0.00541046, 0.0124926)
(0.276411, 0.331385) +- (0.00502445, 0.0109452)
(0.301211, 0.416494) +- (0.00473804, 0.010465)
(0.345596, 0.615919) +- (0.00477823, 0.00649448)
(0.362728, 0.681164) +- (0.00498939, 0.00507105)
(0.383287, 0.744459) +- (0.00486709, 0.0041913)
(0.403796, 0.789404) +- (0.0046782, 0.00339992)
(0.418844, 0.819021) +- (0.0045363, 0.00294163)
};
\addlegendentry{proposed algorithm, 320 units}
\addplot[solid]
coordinates {
(0.156029, 0.0859756) +- (0.00508507, 0.0133354)
(0.245737, 0.164855) +- (0.00482973, 0.0126054)
(0.296806, 0.28165) +- (0.00492433, 0.0122639)
(0.321653, 0.368871) +- (0.00455477, 0.0111873)
(0.36933, 0.58327) +- (0.00514214, 0.00652194)
(0.387559, 0.654551) +- (0.00519268, 0.00504621)
(0.409804, 0.722321) +- (0.0048596, 0.00425663)
(0.428943, 0.773109) +- (0.00481172, 0.00356555)
(0.443752, 0.802831) +- (0.00508116, 0.00321608)
};
\addlegendentry{proposed algorithm, 640 units}
\addplot[thick, dotted]
coordinates {
(0.137347, 0.187314) +- (0.00417829, 0.0140995)
(0.209375, 0.327001) +- (0.00506164, 0.0121331)
(0.246112, 0.451061) +- (0.00501929, 0.0120381)
(0.268306, 0.511717) +- (0.00549038, 0.00954461)
(0.314408, 0.674743) +- (0.0049428, 0.00609106)
(0.334647, 0.736792) +- (0.00497782, 0.00440475)
(0.351075, 0.784691) +- (0.00485923, 0.00409562)
(0.370752, 0.825903) +- (0.00458584, 0.00319301)
(0.386705, 0.848442) +- (0.00466367, 0.00276226)
};
\addlegendentry{edge-exclusion, 160 units}
\addplot[thick, dashed]
coordinates {
(0.143737, 0.113036) +- (0.00423325, 0.0136753)
(0.239671, 0.261068) +- (0.00541046, 0.0128498)
(0.276411, 0.367107) +- (0.00502445, 0.0108611)
(0.301211, 0.452274) +- (0.00473804, 0.0101092)
(0.345596, 0.643941) +- (0.00477823, 0.00616115)
(0.362728, 0.704371) +- (0.00498939, 0.00479211)
(0.383287, 0.762524) +- (0.00486709, 0.00394751)
(0.403796, 0.803292) +- (0.0046782, 0.00317717)
(0.418844, 0.830323) +- (0.0045363, 0.00278467)
};
\addlegendentry{edge-exclusion, 320 units}
\addplot[thick, solid]
coordinates {
(0.156029, 0.0926945) +- (0.00508507, 0.0132183)
(0.245737, 0.188481) +- (0.00482973, 0.0132082)
(0.296806, 0.317215) +- (0.00492433, 0.012399)
(0.321653, 0.407084) +- (0.00455477, 0.0108796)
(0.36933, 0.614345) +- (0.00514214, 0.00634231)
(0.387559, 0.680836) +- (0.00519268, 0.00470975)
(0.409804, 0.742618) +- (0.0048596, 0.00401482)
(0.428943, 0.78859) +- (0.00481172, 0.00340636)
(0.443752, 0.815625) +- (0.00508116, 0.00303817)
};
\addlegendentry{edge-exclusion, 640 units}
\end{semilogyaxis}
\end{tikzpicture}

%% file: Definitions/bbp_10p_50.tex
\begin{tikzpicture}
\begin{semilogyaxis}[xlabel = {network utilization}, ylabel = {bandwidth block-\\ing probability}, height = 6.5 cm, width = 8 cm, grid = both, ylabel style = {align = center, xshift = -0.2 cm}, legend columns = 3, legend to name = bbp, legend style = {/tikz/every even column/.append style = {column sep = 0.25 cm}}, xmin = 0.09, xmax = 0.4, ymin = 0.07, ymax = 1, ]
\addplot[dotted]
coordinates {
(0.117661, 0.187918) +- (0.00569911, 0.0195756)
(0.188487, 0.348916) +- (0.00599344, 0.0166786)
(0.220736, 0.454296) +- (0.0059305, 0.0134528)
(0.247075, 0.528205) +- (0.00565563, 0.0123165)
(0.291796, 0.692383) +- (0.00524972, 0.00689472)
(0.31307, 0.743572) +- (0.00579297, 0.00508437)
(0.332693, 0.784671) +- (0.00505063, 0.00460328)
(0.351376, 0.822168) +- (0.00446423, 0.00373271)
(0.362726, 0.846413) +- (0.00483861, 0.00298562)
};
\addlegendentry{proposed algorithm, 160 units}
\addplot[dashed]
coordinates {
(0.102604, 0.156743) +- (0.00513301, 0.0166416)
(0.181763, 0.327227) +- (0.00564969, 0.0152491)
(0.226664, 0.445445) +- (0.00565136, 0.0128068)
(0.244222, 0.513844) +- (0.00534454, 0.0117268)
(0.297387, 0.690631) +- (0.004939, 0.0065155)
(0.31379, 0.740104) +- (0.00565221, 0.00542711)
(0.334351, 0.784644) +- (0.00489876, 0.00461109)
(0.350942, 0.818483) +- (0.00452578, 0.00399995)
(0.364262, 0.842974) +- (0.00494357, 0.0031526)
};
\addlegendentry{proposed algorithm, 320 units}
\addplot[solid]
coordinates {
(0.0966696, 0.132758) +- (0.00420626, 0.0160736)
(0.188585, 0.332381) +- (0.00615896, 0.0159343)
(0.228492, 0.450887) +- (0.005742, 0.0134491)
(0.249141, 0.515286) +- (0.00559674, 0.0123633)
(0.293901, 0.676593) +- (0.00538317, 0.00714514)
(0.3112, 0.730607) +- (0.00549271, 0.00581541)
(0.331432, 0.780947) +- (0.00564668, 0.00429715)
(0.349545, 0.817473) +- (0.00518425, 0.00361318)
(0.364471, 0.841662) +- (0.00502601, 0.00305581)
};
\addlegendentry{proposed algorithm, 640 units}
\addplot[thick, dotted]
coordinates {
(0.117661, 0.205481) +- (0.00569911, 0.0200223)
(0.188487, 0.374579) +- (0.00599344, 0.0171135)
(0.220736, 0.478761) +- (0.0059305, 0.0131966)
(0.247075, 0.556683) +- (0.00565563, 0.0123698)
(0.291796, 0.713769) +- (0.00524972, 0.00681416)
(0.31307, 0.760716) +- (0.00579297, 0.00490829)
(0.332693, 0.798677) +- (0.00505063, 0.00441582)
(0.351376, 0.833471) +- (0.00446423, 0.00354404)
(0.362726, 0.85533) +- (0.00483861, 0.00282912)
};
\addlegendentry{edge-exclusion, 160 units}
\addplot[thick, dashed]
coordinates {
(0.102604, 0.167734) +- (0.00513301, 0.0171261)
(0.181763, 0.349845) +- (0.00564969, 0.0154039)
(0.226664, 0.474011) +- (0.00565136, 0.0127066)
(0.244222, 0.541524) +- (0.00534454, 0.0114339)
(0.297387, 0.712026) +- (0.004939, 0.00632214)
(0.31379, 0.757993) +- (0.00565221, 0.00526582)
(0.334351, 0.798782) +- (0.00489876, 0.00445116)
(0.350942, 0.828969) +- (0.00452578, 0.00380551)
(0.364262, 0.852244) +- (0.00494357, 0.00303098)
};
\addlegendentry{edge-exclusion, 320 units}
\addplot[thick, solid]
coordinates {
(0.0966696, 0.140139) +- (0.00420626, 0.0165786)
(0.188585, 0.356027) +- (0.00615896, 0.0170421)
(0.228492, 0.47863) +- (0.005742, 0.0133788)
(0.249141, 0.544488) +- (0.00559674, 0.0123533)
(0.293901, 0.700154) +- (0.00538317, 0.00690786)
(0.3112, 0.749674) +- (0.00549271, 0.00551358)
(0.331432, 0.795424) +- (0.00564668, 0.00395703)
(0.349545, 0.828428) +- (0.00518425, 0.00340511)
(0.364471, 0.850492) +- (0.00502601, 0.00281782)
};
\addlegendentry{edge-exclusion, 640 units}
\end{semilogyaxis}
\end{tikzpicture}

%% file: Definitions/bbp_10_100.tex
\begin{tikzpicture}
\begin{semilogyaxis}[xlabel = {network utilization}, ylabel = {bandwidth block-\\ing probability}, height = 6.5 cm, width = 8 cm, grid = both, ylabel style = {align = center, xshift = -0.2 cm}, legend columns = 3, legend to name = bbp, legend style = {/tikz/every even column/.append style = {column sep = 0.25 cm}}, xmin = 0.1, xmax = 0.5, ymin = 0.04, ymax = 1, ]
\addplot[dotted]
coordinates {
(0.138656, 0.116051) +- (0.00349105, 0.00914866)
(0.217396, 0.286257) +- (0.00338113, 0.0094087)
(0.250139, 0.383975) +- (0.00342693, 0.00894517)
(0.265751, 0.461173) +- (0.00358388, 0.00871232)
(0.304259, 0.645771) +- (0.00306796, 0.00463373)
(0.314785, 0.701414) +- (0.00331254, 0.00386843)
(0.333959, 0.758602) +- (0.00298775, 0.00289875)
(0.350223, 0.80071) +- (0.00274093, 0.00249483)
(0.361814, 0.825631) +- (0.00292915, 0.00191617)
};
\addlegendentry{proposed algorithm, 160 units}
\addplot[dashed]
coordinates {
(0.158062, 0.0688973) +- (0.00276868, 0.00686245)
(0.2369, 0.18403) +- (0.00294746, 0.0089178)
(0.271127, 0.305229) +- (0.00297365, 0.00865321)
(0.291183, 0.406675) +- (0.00290559, 0.00701709)
(0.327215, 0.605199) +- (0.00306662, 0.00471769)
(0.341482, 0.668359) +- (0.0030209, 0.00374815)
(0.358734, 0.729757) +- (0.00307359, 0.00277564)
(0.375301, 0.77769) +- (0.00269418, 0.00234887)
(0.386229, 0.807141) +- (0.00275794, 0.002054)
};
\addlegendentry{proposed algorithm, 320 units}
\addplot[solid]
coordinates {
(0.148295, 0.0459269) +- (0.00295821, 0.00589434)
(0.238104, 0.117574) +- (0.00267535, 0.00768948)
(0.284313, 0.243804) +- (0.00252097, 0.00830516)
(0.307893, 0.35358) +- (0.00301953, 0.00690495)
(0.347749, 0.571056) +- (0.00321284, 0.00396929)
(0.363006, 0.640899) +- (0.00312716, 0.00350564)
(0.380794, 0.708732) +- (0.00298298, 0.00270283)
(0.397339, 0.75988) +- (0.0029367, 0.00227934)
(0.409587, 0.790032) +- (0.00287135, 0.002045)
};
\addlegendentry{proposed algorithm, 640 units}
\addplot[thick, dotted]
coordinates {
(0.138656, 0.131932) +- (0.00349105, 0.00967737)
(0.217396, 0.318951) +- (0.00338113, 0.0100431)
(0.250139, 0.424061) +- (0.00342693, 0.00937779)
(0.265751, 0.499634) +- (0.00358388, 0.00882828)
(0.304259, 0.676209) +- (0.00306796, 0.00446653)
(0.314785, 0.727514) +- (0.00331254, 0.00368641)
(0.333959, 0.780879) +- (0.00298775, 0.00273563)
(0.350223, 0.818197) +- (0.00274093, 0.00233948)
(0.361814, 0.840652) +- (0.00292915, 0.00178703)
};
\addlegendentry{edge-exclusion, 160 units}
\addplot[thick, dashed]
coordinates {
(0.158062, 0.0800086) +- (0.00276868, 0.00733575)
(0.2369, 0.216886) +- (0.00294746, 0.0103231)
(0.271127, 0.347153) +- (0.00297365, 0.00925994)
(0.291183, 0.44868) +- (0.00290559, 0.00713667)
(0.327215, 0.640508) +- (0.00306662, 0.0045763)
(0.341482, 0.698062) +- (0.0030209, 0.00358045)
(0.358734, 0.753993) +- (0.00307359, 0.00269101)
(0.375301, 0.797533) +- (0.00269418, 0.00220706)
(0.386229, 0.824175) +- (0.00275794, 0.00193923)
};
\addlegendentry{edge-exclusion, 320 units}
\addplot[thick, solid]
coordinates {
(0.148295, 0.048834) +- (0.00295821, 0.00586906)
(0.238104, 0.141959) +- (0.00267535, 0.0087936)
(0.284313, 0.284009) +- (0.00252097, 0.00911841)
(0.307893, 0.3975) +- (0.00301953, 0.00718148)
(0.347749, 0.608622) +- (0.00321284, 0.0039079)
(0.363006, 0.673145) +- (0.00312716, 0.00336003)
(0.380794, 0.735356) +- (0.00298298, 0.00259409)
(0.397339, 0.781548) +- (0.0029367, 0.00217112)
(0.409587, 0.808716) +- (0.00287135, 0.00191067)
};
\addlegendentry{edge-exclusion, 640 units}
\end{semilogyaxis}
\end{tikzpicture}

%% file: Definitions/bbp_10p_100.tex
\begin{tikzpicture}
\begin{semilogyaxis}[xlabel = {network utilization}, ylabel = {bandwidth block-\\ing probability}, height = 6.5 cm, width = 8 cm, grid = both, ylabel style = {align = center, xshift = -0.2 cm}, legend columns = 3, legend to name = bbp, legend style = {/tikz/every even column/.append style = {column sep = 0.25 cm}}, xmin = 0.09, xmax = 0.4, ymin = 0.04, ymax = 1, ]
\addplot[dotted]
coordinates {
(0.119977, 0.134538) +- (0.00348355, 0.0122781)
(0.190231, 0.306209) +- (0.0039338, 0.0117911)
(0.226321, 0.430678) +- (0.00411252, 0.0107094)
(0.243865, 0.511778) +- (0.00379688, 0.00716331)
(0.288175, 0.678367) +- (0.00338089, 0.00508991)
(0.299012, 0.719157) +- (0.00349773, 0.0042853)
(0.316203, 0.768304) +- (0.00346146, 0.00306236)
(0.329426, 0.807311) +- (0.00298566, 0.00257395)
(0.341588, 0.833822) +- (0.00293003, 0.00206664)
};
\addlegendentry{proposed algorithm, 160 units}
\addplot[dashed]
coordinates {
(0.11443, 0.118504) +- (0.00335236, 0.0105291)
(0.188332, 0.28926) +- (0.0038314, 0.0143945)
(0.226132, 0.41634) +- (0.00412051, 0.0108044)
(0.248117, 0.498683) +- (0.00396907, 0.00947891)
(0.288263, 0.670227) +- (0.00365752, 0.00525658)
(0.301603, 0.717848) +- (0.0034388, 0.00435077)
(0.315909, 0.763469) +- (0.00315253, 0.0029817)
(0.330948, 0.804467) +- (0.00306049, 0.0025916)
(0.342636, 0.829368) +- (0.00289001, 0.00199304)
};
\addlegendentry{proposed algorithm, 320 units}
\addplot[solid]
coordinates {
(0.103775, 0.0915734) +- (0.00298092, 0.0107483)
(0.197487, 0.30844) +- (0.00356882, 0.0117932)
(0.231169, 0.422146) +- (0.00336791, 0.0107633)
(0.247417, 0.493602) +- (0.0037526, 0.0091739)
(0.288076, 0.665034) +- (0.00345113, 0.00604627)
(0.303308, 0.712799) +- (0.00356083, 0.00440595)
(0.31731, 0.759222) +- (0.00361482, 0.00348182)
(0.331111, 0.80174) +- (0.00330166, 0.00260654)
(0.341973, 0.828667) +- (0.00287774, 0.00222391)
};
\addlegendentry{proposed algorithm, 640 units}
\addplot[thick, dotted]
coordinates {
(0.119977, 0.149393) +- (0.00348355, 0.0130305)
(0.190231, 0.335554) +- (0.0039338, 0.0124317)
(0.226321, 0.465105) +- (0.00411252, 0.0111208)
(0.243865, 0.547811) +- (0.00379688, 0.00712489)
(0.288175, 0.706519) +- (0.00338089, 0.0047319)
(0.299012, 0.743385) +- (0.00349773, 0.00414109)
(0.316203, 0.788373) +- (0.00346146, 0.00300594)
(0.329426, 0.823704) +- (0.00298566, 0.00248271)
(0.341588, 0.847884) +- (0.00293003, 0.00205335)
};
\addlegendentry{edge-exclusion, 160 units}
\addplot[thick, dashed]
coordinates {
(0.11443, 0.132126) +- (0.00335236, 0.0110483)
(0.188332, 0.31941) +- (0.0038314, 0.0149504)
(0.226132, 0.450874) +- (0.00412051, 0.011361)
(0.248117, 0.533994) +- (0.00396907, 0.0098582)
(0.288263, 0.699427) +- (0.00365752, 0.00529289)
(0.301603, 0.743684) +- (0.0034388, 0.00402865)
(0.315909, 0.784238) +- (0.00315253, 0.00292414)
(0.330948, 0.821281) +- (0.00306049, 0.00253702)
(0.342636, 0.843892) +- (0.00289001, 0.00187169)
};
\addlegendentry{edge-exclusion, 320 units}
\addplot[thick, solid]
coordinates {
(0.103775, 0.102733) +- (0.00298092, 0.0122089)
(0.197487, 0.340526) +- (0.00356882, 0.0124532)
(0.231169, 0.458293) +- (0.00336791, 0.0110309)
(0.247417, 0.529542) +- (0.0037526, 0.00914936)
(0.288076, 0.694112) +- (0.00345113, 0.00588073)
(0.303308, 0.738096) +- (0.00356083, 0.00414435)
(0.31731, 0.780413) +- (0.00361482, 0.00331117)
(0.331111, 0.81902) +- (0.00330166, 0.0024783)
(0.341973, 0.843587) +- (0.00287774, 0.00202216)
};
\addlegendentry{edge-exclusion, 640 units}
\end{semilogyaxis}
\end{tikzpicture}